\documentclass{revtex4}

\usepackage{epsf}
\usepackage[dvips]{color}

\newcommand\Ro{\mbox{\textit{Ro}}}  

\newcommand{\f}[2]{\frac{#1}{#2}}
\newcommand{\paf}[2]{\ensuremath{\left(\f{#1}{#2}\right)}}

\newcommand{\Int}[2]{\ensuremath{\mathchoice%
	{\displaystyle\int_{#1}^{#2}}
	{\displaystyle\int_{#1}^{#2}}
	{\int_{#1}^{#2}}
	{\int_{#1}^{#2}}
	}}

\newcommand{\dd}{d}

\newcommand{\Dp}[2]{\ensuremath{\f{\partial#1}{\partial#2}}} 
\newcommand{\Dpt}[1]{\ensuremath{\Dp{#1}{t}}}
\newcommand{\DDp}[2]{\ensuremath{\f{\partial^2#1}{\partial{#2}^2}}}

\newcommand{\mean}[1]{\ensuremath{\left\langle#1\right\rangle}}
\newcommand{\avg}[1]{\ensuremath{\overline{#1}}}
\newcommand{\pa}[1]{\ensuremath{\left(#1\right)}}
\newcommand{\abs}[1]{\ensuremath{\left|#1\right|}}

\newcommand{\rot}{\ensuremath{\vect{\nabla}\vectoriel}}
\newcommand{\grad}{\ensuremath{\vect{\nabla}}}
\newcommand{\lap}{\ensuremath{\nabla^2}}

\newcommand{\vect}[1]{\ensuremath{%
        {\mbox{\mathversion{bold}\ensuremath{#1}}}%
        }}

\newcommand{\U}[1]{\ensuremath{\mathrm{~#1}}}

\newcommand{\vectoriel}{\ensuremath{\wedge}}

\newcommand{\PT}{Proudman-Taylor }

\begin{document}
\draft
\title{Quasi-geostrophic model of the instabilities of the Stewartson layer}
%
\author{Nathana\"el Schaeffer} 
\affiliation{LGIT, Universit\'e Joseph Fourier,
Grenoble, France}
\email[]{Nathanael.Schaeffer@ujf-grenoble.fr}

\author{Philippe Cardin}
\affiliation{LGIT, Universit\'e Joseph Fourier and CNRS,
Grenoble, France}
\date{\today}

\begin{abstract}
We study the destabilization of a shear layer, produced by differential
rotation of a rotating axisymmetric container.
For small forcing, this produces a shear layer, which has been
studied by Stewartson and is almost invariant along the rotation axis.
When the forcing increases, instabilities develop.
To study the asymptotic regime (very low Ekman number $E$), we develop a quasi-geostrophic two-dimensional model, whose main original feature is to handle the mass conservation correctly, resulting in a divergent two-dimensional flow, and valid for any container provided that the top and bottom have finite slopes.
We use it to derive scalings and asymptotic laws by a simple linear theory, extending
the previous analyses to large slopes (as in a sphere), for which we find different scaling laws. For a flat container, the critical Rossby number for the onset of instability evolves as $E^{3/4}$ and may be understood as a Kelvin-Helmoltz shear instability. For a sloping container, the instability is a Rossby wave with a critical Rossby number proportional to $\beta E^{1/2}$, where $\beta$ is related to the slope.
We also investigate the asymmetry between positive and negative differential rotation and
propose corrections for finite Ekman and Rossby numbers.
Implemented in a numerical code, our model allows us to study the onset over a broad range of parameters, determining the threshold but also other features such as the spatial structure.
We also present a few experimental results, validating our model and showing its limits.

\end{abstract}

\maketitle

\section{Introduction}


\subsection{Background}

The destabilization of a shear layer in a rotating system is
of general geophysical interest. Shear layers dominated by the Coriolis force,
are found in natural systems, as in the atmosphere or in the ocean, as well as on other planets like the jets of Jupiter, which might be unstable and responsible for the observed big eddies.

This work considers the shear layer produced by differential rotation of the boundaries
of a fast rotating container, in axisymmetric geometries.
The linear problem resulting in cylindrical shear layers aligned with the axis
of rotation has been studied by \cite{Stew:sphere} in the spherical shell geometry.
He pointed out the nested shear layer structure and their scaling.
This problem is closely related to the simpler case of a split
cylinder solved earlier by \cite{Stew:cyl} too. In that case, the structure of the
shear layer is simpler, but there is still a nested structure.
The general structure of these detached shear layers is studied by \cite{Moore}
by focusing on the singularities of the boundary-layer equations.
Later, a numerical study has been done by \cite{Dormy:MHD}, recovering the features
and scalings predicted by \citeauthor{Stew:sphere}.

The linear stability of this kind of layers has been theoretically
investigated by \cite{Busse:68} who applied a linear stability theory
to the pressure equation in both cases of constant and varying depth.
He neglected the bulk viscosity but kept the Ekman friction, recovering some kind of Rayleigh criterion, showing that for the flat case, one should expect a critical Rossby number $\Ro_c \sim E^{3/4}$ (where $E$ is the Ekman number).
For a varying depth container, he discussed the criterion obtained by \cite{Kuo:49}.

One may notice that the difference between flat and varying depth containers has been made explicit by \cite{Busse:70} for thermal instabilities in rapidly rotating systems : for a flat container the instability is of B\'enard type (critical Rayleigh number $Ra_c$ independent of Ekman number), whereas for a spherical shell the instability is a thermal Rossby wave ($Ra_c \sim E^{-1/3}$).



The setup of \cite{Hide}, with a central horizontal disk rotating differentially in
a flat cylindrical tank, gives unexpected results which are not yet well understood.
They also observed a strong asymmetry between positive and negative differential
rotation.
Later, \cite{Niino} performed a similar experiment but with the differential rotation
concentrated on the bottom of their flat tank.
They find a critical local Reynolds number for the destabilization of the flow, independent of the Ekman number.
Their linear theory which includes both Ekman friction and bulk viscosity is in very good agreement with their results. They point out that the bulk
viscosity is very important to reproduce accurately the experimental results.
They showed that a critical local Reynolds number (defined for the shear layer) describes the threshold of instability properly.

Later, \cite{Read} reproduced almost exactly the case studied by
\cite{Stew:cyl}. The stability threshold they found is still governed by a critical local Reynolds number, but the theory of \citeauthor{Niino} predicts a value much lower than the
observed one. They also observed that the behavior of the flow does not depend strongly on the sign of the differential rotation.

The shear-flow instabilities in a parabolic tank and a free surface was investigated
by \cite{Nielsen}. They studied both cases with constant depth or
varying depth and pointed out that the varying depth has a stabilizing effect,
as predicted by the linear theory.
However, their setup uses the centrifugal force to adjust the slope of the free surface, so that the slope is directly related to the rotation rate.

More recently, using a 3D code, \cite{Hollerbach:1} studied the differences between positive and negative differential rotation for the onset of these instabilities, pointing out the importance of the geometry of the container.

\begin{figure}
  \centerline{ \epsfxsize 8cm \epsffile{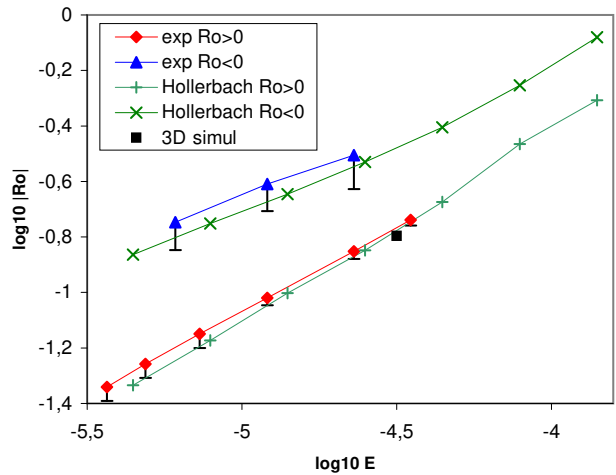}}
	\caption{Critical Rossby number versus Ekman number for the spherical shell geometry.
	The results obtained numerically by \cite{Hollerbach:1} and our lab experiment are in good agreement. The symbols represent established instability, and the error bars the domain where instability was not always clear. "3D simul" is a run with our own 3D code.
}
	\label{fig:seuil_sphere}
\end{figure}

\cite{Hollerbach:1} computed numerical solutions for the instability of the Stewartson layer between two rotating spheres. Meanwhile, we ran the corresponding experiment, measuring the threshold. Details of the experimental setup can be found in \S\ref{sec:ExpSetup}.
The results of theses two studies are in agreement, and summarized in figure \ref{fig:seuil_sphere}.
The study of \cite{Dormy:MHD} showed that the asymptotic regime of the Stewartson
problem cannot be illustrated numerically for Ekman numbers higher than $10^{-5}$.
Unfortunately, neither the experimental nor the numerical approaches were able to reach Ekman numbers lower than $10^{-5.5}$. This number should be compared with geophysical flows, where Ekman numbers can be much smaller : for the Jovian atmosphere, the oceans and the Earth core the Ekman number is smaller than $10^{-10}$.

\subsection{Outline}

To study the very low Ekman number regimes we use a quasi-geostrophic (QG) model, detailed in \S\ref{sec:2Dmodel} :
we derive the equations of an improved QG-model, that will be used to address the problem in the next sections, through numerical computations and theoretical analysis of the equations. This QG-model is improved over previous ones by taking into account the mass conservation and Ekman friction. The latter is also dynamically important for non-linear regimes because of its dissipative nature.
We were able to benchmark this model with some fully 3D computations by \cite{Hollerbach:1} and the results are quite encouraging.

\S\ref{sec:StewLayer} presents results concerning the Stewartson layer,
comparing the QG-model with existing results, showing its ability to reproduce the axisymmetric flow.
A numerical code based on the quasi-geostrophic approximation allows us to reach Ekman numbers as low as $10^{-10}$ which is very likely to lie in the asymptotic regime, from which extrapolations can be made to even lower Ekman numbers.

In \S\ref{sec:SpatStruc}, using numerical calculations as well as theoretical arguments, we investigate the spatial structure of the instability and its asymptotic evolution when lowering the Ekman number. When a $\beta$-effect is present, the instability is a Rossby wave, and consequently we find that its spatial structure changes dramatically with the sign of the differential rotation. This explains the asymmetric sensitivity to geometry studied by \cite{Hollerbach:1}.
Another interesting feature is that the radial extent of the instability may be independent of the Ekman number in some particular cases.

In \S\ref{sec:Asymptotic} we derive the scaling laws expected in the asymptotic regime, focusing on the stability threshold.
For a flat container, an $E^{3/4}$ law is recovered for the critical Rossby number.
We extend this study to finite slopes, in which case the scaling changes to $\beta E^{1/2}$, where $\beta$ is related to the slope. This scaling is the one that should be applied to spherical shells.
We perform an analysis leading to a Rayleigh-like criterion, which allows us to evaluate quantitatively the effect of the asymmetry induced by the Rossby wave mechanism.
Numerical results obtained with our QG-model are also reported here, for various geometries.

\S\ref{sec:Exp} compares our results with previous experiments, as well as with our own, to span a wide range of different geometries. Despite the lack of quantitative agreement between numerical results obtained with the QG-model and the experimental data, the global behaviour of the instability is captured.
Finally we conclude in \S\ref{sec:ccl} with a discussion.

\section{Quasi-geostrophic model}
\label{sec:2Dmodel}

The two-dimensional quasi-geostrophic (QG) model we present here has been developed
to study the dynamics of a fluid in a rapidly rotating container,
invariant by rotation around the rotation axis, without being restricted to the
small slope case.
We use the cylindrical coordinate system ($\vect{e_r}, \vect{e_{\phi}}, \vect{e_z}$),
with $r$ the distance to the axis, $\phi$
the angle in the plane perpendicular to that axis, and $z$ the height.
The container is defined by
two functions $z=L_+(r)$ and $z=L_-(r)$, for the top and the bottom surfaces
respectively.
Here, we first restrict to the simpler case where $L(r) = L_+(r)  = - L_-(r)$.


In a frame rotating at angular velocity $\Omega_0$ around the vertical axis $z$,
the flow is described by the Navier-Stokes equation for an incompressible 
fluid with constant density, including the Coriolis force :
\begin{equation}
\label{eq:NS}
	\f{\partial \vect{u}}{\partial t} +
		\pa{\vect{u}.\grad}\vect{u} + 2\vect{\Omega_0} \vectoriel \vect{u}
		= -\grad \Pi + \nu \lap \vect{u}
\end{equation}
where $\vect{u}$ is the fluid velocity field, $\nu$ the kinematic
viscosity, and $\Pi$ the reduced pressure field, including centrifugal
and gravity potentials.
We introduce a velocity scale $U$, a time scale ${\Omega_0}^{-1}$
and as typical length scale $R$ the radial size of the container.
We then rewrite equation \ref{eq:NS} with non-dimensional quantities :
\begin{equation}
\label{eq:NSAdim}
	\f{\partial \vect{u}}{\partial t} +
		\Ro \pa{\vect{u}.\grad}\vect{u} + 2\vect{e_z} \vectoriel \vect{u}
		= -\grad \Pi + E \lap \vect{u}
\end{equation}
with $\Ro \equiv U/R\Omega_0$ the Rossby number,
$E \equiv \nu/R^2\Omega_0$ the Ekman number,
and $\vect{e_z}$ the unit vector along the rotation axis.

\noindent In our study, we take for typical velocity the one based on
the differential rotation
$U = R\Delta\Omega$, which is supposed to dominate. Hence, the
Rossby number reduces to $\Delta\Omega/\Omega_0$.

\subsection{$z$-averaged vorticity equation}

For vanishing Ekman and Rossby number, the \PT theorem ensures that the flow
is invariant along the rotation axis direction ($\Dp{\vect{u}}{z} = 0$).
This suggests that for small Rossby and Ekman numbers, the $z$-variations
should be weak and we expect the flow to be mainly two-dimensional for
long-period motions ($ \gg \Omega_0^{-1}$).

We define the $z$-averaging $\mean{\;}$ by
\begin{equation}
  \mean{u} = \f{1}{2L} \Int{-L}{+L} u(r,z) \: \dd z
\end{equation}
Keeping in mind that $L$ depends on $r$, we have
$$
  \Dp{\mean{u}}{r} = \mean{\Dp{u}{r}} + 
    \f{1}{L}\Dp{L}{r} \pa{ \f{u(r,+L) + u(r,-L)}{2} - \mean{u(r,z)} }
$$
Hence, a sufficient condition for the $r$-derivative
to commute with $z$-averaging is that $u$ is independent of $z$,
that is $u = \mean{u}$.

We now average the $z$-component of the curl of equation \ref{eq:NSAdim}
with the additional constraint that $u_r$ and $u_{\phi}$ are independent
of $z$ (so that we can permute $r$-derivation and $z$-averaging).
The $z$-component of the vorticity $\vect{\omega}$ is denoted by $\omega$
and is also $z$-invariant.
We obtain
\begin{equation}
\label{eq:Vort2D}
        \Dpt{\omega} + \Ro \pa{ u_r\Dp{\omega}{r} + 
		\f{u_{\phi}}{r} \Dp{\omega}{\phi} }
        - (2+\Ro\:\omega) \mean{\Dp{u_z}{z}} = E \lap \omega
\end{equation}
Note that we keep the vortex stretching term $\mean{\Dp{u_z}{z}}$.
All other terms vanish, without any condition on $u_z$.

\subsection{Mass conservation}

The local mass conservation equation for an incompressible fluid,
\begin{equation}
\label{eq:Mass}
\f{1}{r}\Dp{r u_r}{r}
        + \f{1}{r}\Dp{u_{\phi}}{\phi} + \Dp{u_z}{z} = 0
\end{equation}
implies that $\Dp{u_z}{z}$ is independent of $z$, so that we can
now draw a global picture of the flow in the frame of our QG-model.
\begin{itemize}
  \item $u_r$ and $u_{\phi}$ are $z$-invariant
  \item $u_z$ depends linearly on $z$.
\end{itemize}

\subsection{Boundary conditions and Ekman pumping}
\label{sec:BC}

At the boundaries, Ekman layers will smooth out the vorticity jump.
Their thickness is supposed to be small ($E \ll 1$),
and we will use the asymptotic expression given by \cite{Greenspan:book},
equation 2.6.13 :
\begin{equation}
\label{eq:EkPump}
\left. \vect{u}^1.\vect{n} \right|_{\pm L} = 
\f{1}{2}E^{1/2}\, \vect{n}.\rot
        \paf{\vect{n}\vectoriel\pa{\vect{u}-\vect{U_b}}
        \mp \pa{\vect{u}-\vect{U_b}}}
        {\sqrt{\abs{\vect{n}.\vect{e_z}}}}
\end{equation}
where $\vect{U_b}$ is the velocity of the boundary.
An important point is that expression \ref{eq:EkPump} is valid for
time-dependent flows with time scales larger than a few rotation periods.
Greenspan used that expression to study the spin-up process in a sphere, which is a time-dependent flow with time scale of order $E^{-1/2}$, and \cite{ViscTurb} show experimental data in very good agreement with this linear theory.
It is the suitable approximation for our study,
as we will find typical period of order $E^{-1/4}$ in section \ref{sec:Asymptotic}.
This gives us the appropriate boundary conditions for our problem, allowing us to
express the vertical velocity.
Defining $\beta$
\begin{equation}
\label{eq:beta}
  \beta \equiv \f{1}{L} \left.\Dp{L}{r}\right|_{z=L}
\end{equation}
we have
\begin{equation}
\label{eq:duzPump}
  \Dp{u_z}{z} = E^{1/2}.P(u_r,u_{\phi},r) + \beta u_r
\end{equation}
where $P$ is a function depending on the geometry of the container and
linear with respect to the jump of $u_r$ and $u_{\phi}$ at the boundary
and to their first order derivatives.
See Appendix \ref{sec:EkP} for the explicit expression for a spherical
container.

In the case of a constant-depth container, equation \ref{eq:duzPump} reduces to
\begin{equation}
\label{eq:EkPumpFlat}
  \Dp{u_z}{z} = -\f{E^{1/2}}{2L} \pa{ \omega - \Omega_b }
\end{equation}
where $\Omega_b$ is the vorticity 
at the boundary.


In the case of vanishing Ekman number, the Ekman pumping could be neglected compared
to the $\beta$ term (see \S\ref{tab:stab_fx}).
However, the Ekman friction is a dissipative term whereas the $\beta$-term is not,
it is thus expected to play an important role for the non-linear dynamics of quasi-geostrophic flows.

\medskip


The problem is now reduced to the determination of two
components of the velocity, $u_r$ and $u_{\phi}$, depending only
on $r$ and $\phi$.
We also need to set a boundary condition on the sides of the container,
to be consistent with the modeling of $\mean{\Dp{u_z}{z}}$, we have to use
a no-slip boundary condition $u_r = u_{\phi} = 0$.
At the center ($r=0$) we impose that $u_r$ and $u_{\phi}$ remain finite.

\subsection{Scalar pseudo-stream function}
\label{sec:scal_func}

We introduce a scalar field $\psi$, defined by
\begin{equation}
\label{eq:ur-psi}
        u_r \equiv \f{1}{r} \Dp{\psi}{\phi}
\end{equation}
$u_{\phi}$ is now constrained by the three-dimensional mass
conservation equation \ref{eq:Mass}, so that for \emph{non-axisymmetric} flows,
\begin{eqnarray}
  u_{\phi} & = & -\Dp{\psi}{r} - \beta\psi \\
  \omega & = & -\lap\psi - \beta \pa{\Dp{\psi}{r} + \f{\psi}{r}} -
\f{\dd \beta}{\dd r} \psi
\end{eqnarray}
With these equations, we can see that the 
$\beta$-terms are negligible
for $\beta \lambda \ll 1$, where $\lambda$ is a typical radial length
scale of the velocity field variations. The model then reduces to the
widely used small slope approximation \cite[see e.g.][]{Busse:70, Aubert:G3, Kiss}.
It will be shown later, that at the stability threshold, the typical
length scale is of order $E^{1/4}$, so that asymptotically and at the threshold of instability, the small slope approximation may even be valid in a sphere.
However, increasing the forcing, large scale structures will eventually
appear, with up to order-one scales, so that we cannot neglect the $\beta$-terms
any more in a fully turbulent computation.

Similar approaches have been developed recently.
\cite{Kiss} proposed a formulation for ocean modeling
but without the mass conservation correction.
\cite{Zavala} include the effect of weak topography and Ekman friction.
The advantage of their formulation is that the Ekman pumping is
considered as a source of horizontal divergence. However the simple expression
they used for the Ekman boundary condition implicitly assumes nearly flat topography (small $\beta$).

\subsection{Axisymmetric flow}

For the axisymmetric flow, the scalar function formalism cannot be applied.
As it is simpler to work with the velocity field rather than with the vorticity,
we project the Navier-Stokes equation \ref{eq:NSAdim} on 
$\vect{e_{\phi}}$, noting that 
$\pa{\vect{e_z}\vectoriel \vect{u}}.\vect{e_{\phi}} = u_r$ and
remembering that $\Dp{u_{\phi}}{z}=0$.
Averaging over $\phi$ to keep only the axisymmetric component, we obtain
\begin{equation}
\label{eq:NLbase}
    \Dp{\avg{u}_{\phi}}{t} + \Ro \pa{ \avg{ u_r\Dp{u_{\phi}}{r} } 
        + \f{\avg{ u_{\phi}u_r}}{r}   
 	} + 2\avg{u}_r = E \pa{\lap \avg{u}_{\phi} - \f{\avg{u}_{\phi}}{r^2}}
\end{equation}
where $\avg{u}$ stands for the $\phi$-average of $u$.

The mass conservation equation reduces to
$$
  \f{1}{r}\Dp{r \, \avg{u}_r}{r} + \Dp{\avg{u}_z}{z} = 0
$$
and using equation \ref{eq:duzPump}, we can show that $\avg{u}_r$ is only due
to the Ekman pumping and is directly related to $\avg{u}_{\phi}$ so that
the axisymmetric part is actually one-dimensional :
$\avg{\vect{u}} = \avg{u}_{\phi}(r) \, \vect{e_{\phi}} + O(E^{1/2})$.
However, non-linear interactions of non-axisymmetric modes can produce
axisymmetric flow, so that we keep the terms involving $\avg{u}_r$.
Furthermore, the Ekman pumping flow $\avg{u}_r^1$ of order $E^{1/2}$ cannot be neglected.
Equation \ref{eq:EkPump} reduces to
\begin{equation}
\label{eq:EkPump0}
        \avg{u}_r^1 = E^{1/2}\f{\alpha}{2L} \pa{\avg{u}_{\phi}-U_b}
\end{equation}
with
\begin{equation}
\label{eq:alpha}
		\alpha \equiv \pa{1+\pa{\f{\dd L}{\dd r}}^2}^{1/4}
\end{equation}

\subsection{Numerical implementation}
\label{sec:num}

We wrote a numerical code, that solves the equation of the model
described above using the pseudo-stream function formulation.
We use a finite difference scheme in the radial direction, and a
Fourier expansion in azimuth.
The time integration is done using the semi-implicit Cranck
Nicholson scheme for the linear part, and a simple Adams-Bashforth
scheme for the non-linear terms, which are computed in direct space,
as in \cite{Aubert:G3}.

Our numerical implementation is able to handle various geometries from flat containers to spherical ones.

When compared to fully 3D computations, our QG-model gives quite good results for the basic state flow (see \S\ref{sec:StewLayer}) as well as for the stability threshold (see fig. \ref{fig:seuil_Hol} and \S\ref{sec:exp_ss} for more details).

\section{Stewartson layers}
\label{sec:StewLayer}


The linear study of shear layers due to differential rotation of the walls
has been done in the case of two concentric
co-rotating disks \citep{Stew:cyl}, and in the case of two concentric
differentially rotating spheres \citep{Stew:sphere}.

\begin{figure}
  \centerline{ \epsfxsize 7cm \epsffile{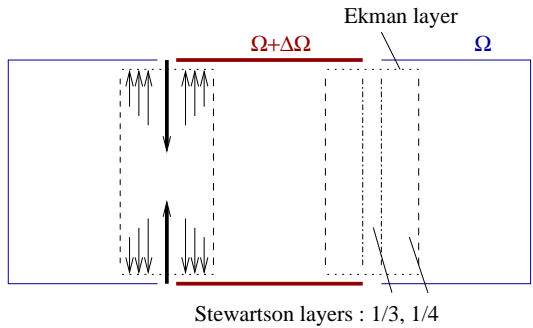}}
	\caption{Sketch of the nested Stewartson layers. The arrows show the Ekman pumping
	inside these layers.}
	\label{fig:stew_layers}
\end{figure}

At first order, the flow is invariant along the direction of the rotation axis,
and consists of a shear layer which is smoothing out the velocity discontinuity.
There are actually two nested layers (see fig. \ref{fig:stew_layers}) with
width scaling like $E^{1/3}$ for the inner layer, and $E^{1/4}$ for the outer
one. Note that the $1/3$ layer is not two-dimensional and smoothes out the
singularity in the Ekman pumping due to the outer layer.
For the spherical shell case, the outer layer width scaling
is slightly modified to $E^{2/7}$ for $r<r_0$ only.
Thus, we can expect that the destabilization of the shear layer in both cases will be closely related.

\subsection{QG-model predictions}

In the linear and stationary regime, when the Rossby number is small
equation \ref{eq:NLbase} reduces to a balance between the Ekman
pumping term \ref{eq:EkPump0} and the viscous forces in the bulk.
\begin{equation}
   \label{eq:baseflow}
   2 E^{1/2}\f{\alpha}{2L} \pa{\avg{u}_{\phi}-U_b} 
	= E \pa{\lap \avg{u}_{\phi} - \f{\avg{u}_{\phi}}{r^2}}
\end{equation}
which defines the Stewartson layer.
Introducing $\Delta$, the width of the Stewartson shear layer we easily recover
the scaling obtained by \citeauthor{Stew:cyl}.
\begin{equation}
		\Delta \sim E^{1/4}   
\end{equation}
Note that the neglected non-linear terms are small if $\Ro /\Delta \ll 1$.
In \S\ref{sec:scaling} we will find that $\Ro \sim E^a$ with $a \ge 1/2$, validating this approximation near the threshold.

One of the major issues with the QG-model, is that it cannot model the inner $E^{1/3}$-layer.
However, according to the numerical study of \cite{Dormy:MHD}, the $E^{1/3}$-layer has a negligible effect on the velocity profile at Ekman numbers smaller than $E=10^{-6}$.

\subsection{Numerical results}

\begin{figure}
  \centerline{ \epsfxsize 8cm \epsffile{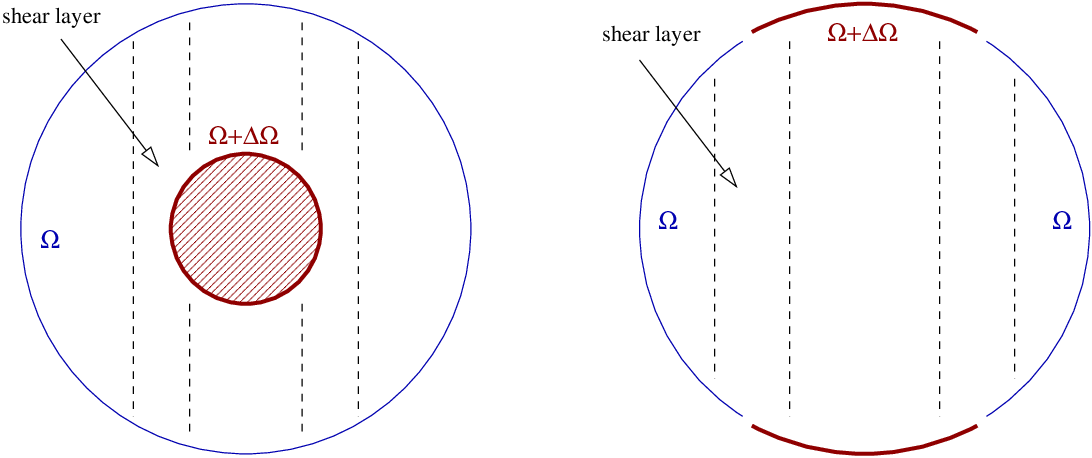}}
	\caption{(a) the spherical shell geometry, studied by
		 \cite{Stew:sphere}.
		 (b) the split-sphere geometry this paper focuses on.}
	\label{fig:sys}
\end{figure}

\begin{figure}
  \centerline{ \epsfxsize 8cm \epsffile{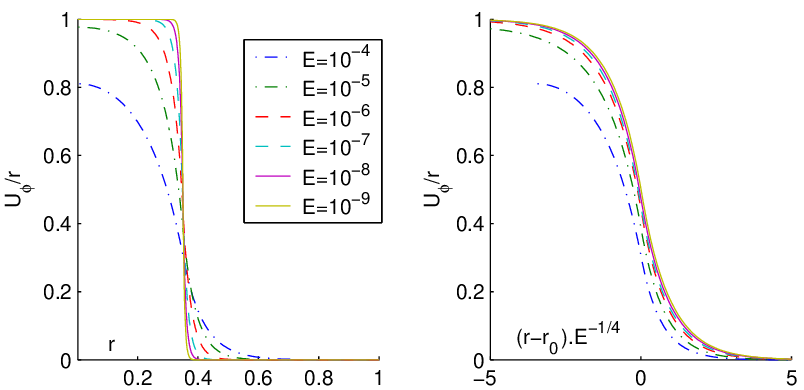}}
	\caption{Basic state flow showing the shear layer which scales with $E^{1/4}$
        in the asymptotic low Ekman number regime, for $r_0 = 0.35$ in
	the split-sphere geometry. (a) shows the raw radial profiles and (b) the same profiles rescaled radially, which merge together when $E$ is sufficiently low ($E \le 10^{-6}$).}
        \label{fig:base_flow}
\end{figure}

The split-sphere geometry consists of a sphere, split at a given cylindrical
radius $r_0$ (see fig. \ref{fig:sys}).
We solve equation \ref{eq:baseflow} in the frame rotating with
the outer part at angular velocity $\Omega_0$,
while the inner part rotates differentially at $\Delta\Omega$
(positive or negative).
This linear problem is solved
by a simple matrix inversion using a finite difference
radial scheme with variable grid spacing.
In this case, the boundary velocity $U_b$ is a simple function,
$U_b=r$ if $r<r_0$ and $U_b=0$ if $r>r_0$.
The resulting shear layer for $r_0=0.35$ is shown
on figure \ref{fig:base_flow}.
Properly scaled, all these profiles can be superposed in the asymptotic, small
Ekman number, regime. The size of the shear layer scales like $E^{1/4}$, as
predicted by \cite{Stew:cyl} and the asymptotic regime seems to be attained
for $E<10^{-6}$.

The spherical shell (see fig. \ref{fig:sys}) geometry is the one
studied originally by \cite{Proudman}, and is also of geophysical relevance because
the core of the Earth is composed of a solid iron ball ($r_0=0.35$) surrounded by liquid iron.
The structure of the Stewartson layer in a spherical shell (including the 
$E^{1/3}$-layer), has been modelled numerically by \cite{Dormy:MHD},
with a fully three dimensional code.
We have compared their results with the ones obtained with our
QG-model for the spherical shell geometry for $\Ro=0$ and find that the difference between the two basic state profiles of the angular velocity is always smaller than $0.03$ for $E=10^{-6}$; and for $E=10^{-8}$ the difference never exceeds $0.008$.
Hence, we may expect the $E^{1/3}$-layer to play only a minor role at such small Ekman numbers.

\section{Spatial structure at the onset}
\label{sec:SpatStruc}

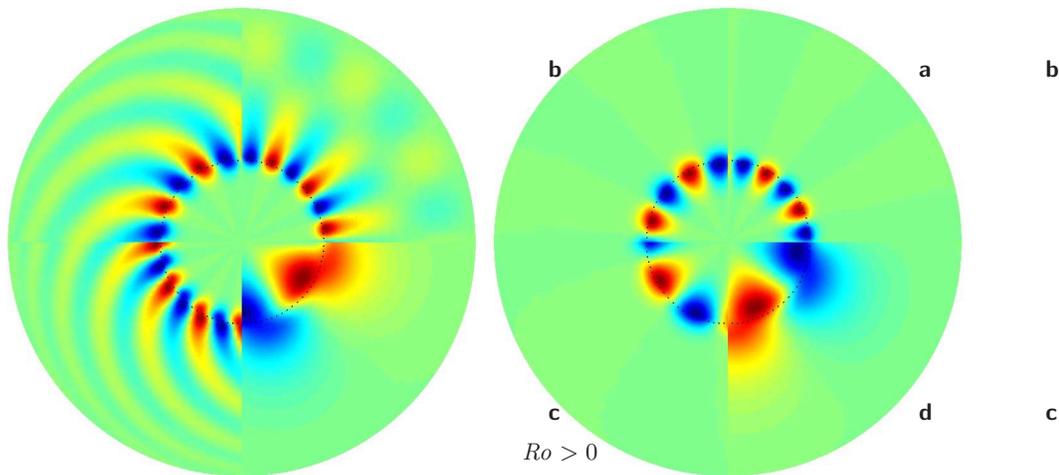
\begin{figure}
  \centerline{  \epsfxsize 6.5cm \epsffile{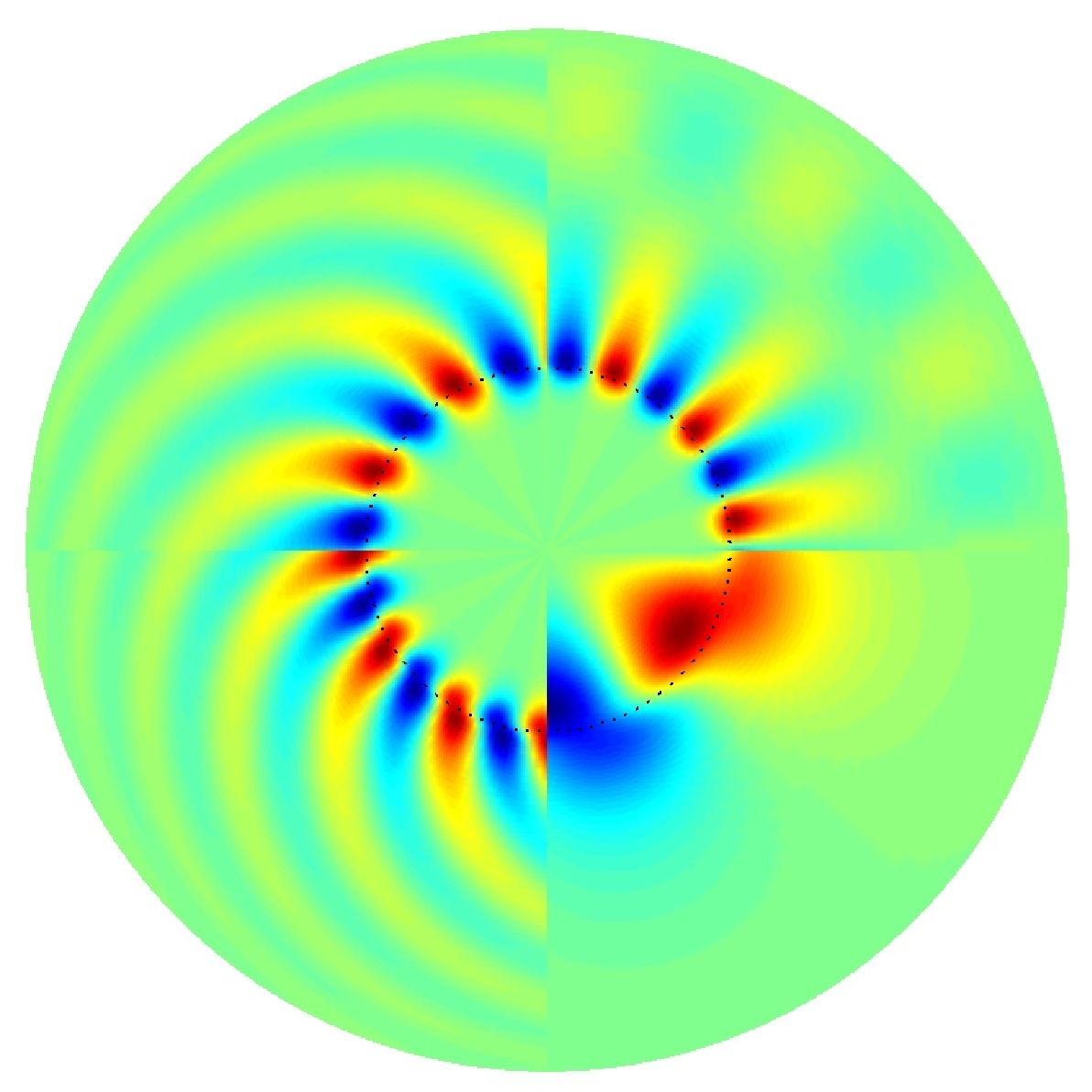}
		\epsfxsize 6.5cm \epsffile{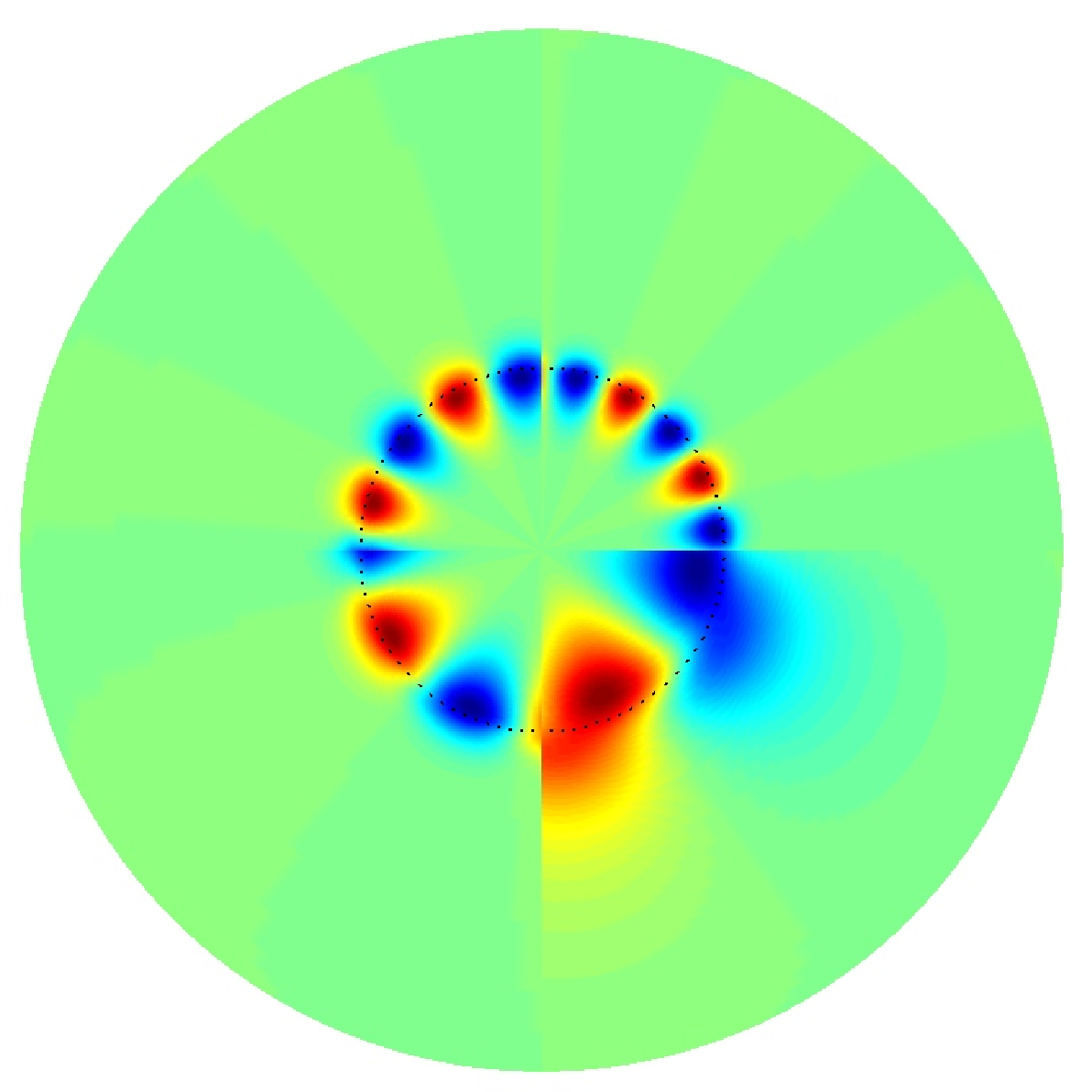}}
  \begin{picture}(0,0)(3,-10)
    \put(20,10){$\Ro>0$}
    \put(340,10){$\Ro<0$}
    \put(170,155){\sffamily \textbf{a}}  \put(170,25){\sffamily \textbf{d}}
    \put(30,155){\sffamily \textbf{b}}  \put(30,25){\sffamily \textbf{c}}
    \put(358,155){\sffamily \textbf{a}}  \put(358,25){\sffamily \textbf{d}}
    \put(218,155){\sffamily \textbf{b}}  \put(218,25){\sffamily \textbf{c}}
  \end{picture}
  \vspace{-5mm}
  \caption{North-pole views of the radial velocity for the first unstable mode at $E=10^{-6}$ for various geometries. Red is positive, blue negative, and green zero.
    The picture on the left is for $\Ro > 0$, and the one on the right
    is for $\Ro < 0$. On each picture, (a) shows the constant $\beta=-1$ geometry; (b) shows the split-sphere case; (c) is the disk setup; and (d) is the unstable mode for a flat container ($\beta=0$).}
	\label{fig:e6crit}
\end{figure}

Using our QG-model, we obtained numerically the first unstable mode 
for four geometries :
the flat container; a container with an exponential shape ($\beta=-1$ everywhere);
the split-sphere geometry (see fig. \ref{fig:sys}); and the disk geometry.
The latter is a split-sphere where the polar caps have been replaced by flat disks,
to mimic our experimental setup (see fig. \ref{fig:exp_setup}).
The first unstable mode is represented on figure \ref{fig:e6crit} for all these geometries.
Note that in the reference frame chosen for our computations and for small enough
Ekman numbers, the mean flow vanishes in the outer region, whereas the inner
region is almost in solid body rotation at the angular velocity $\Delta\Omega = \Ro \Omega_0$
(see figure \ref{fig:base_flow}).

\subsection{Dependence on the sign of the Rossby number for $\beta \neq 0$}
\label{sec:sgnRo}

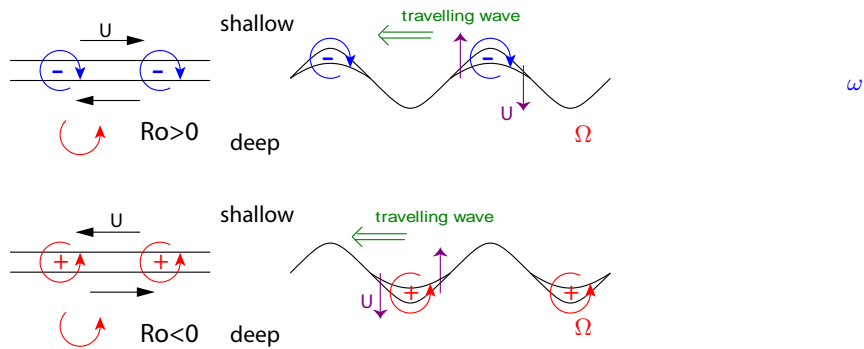
\begin{figure}
  \centerline{ \epsfxsize 8cm \epsffile{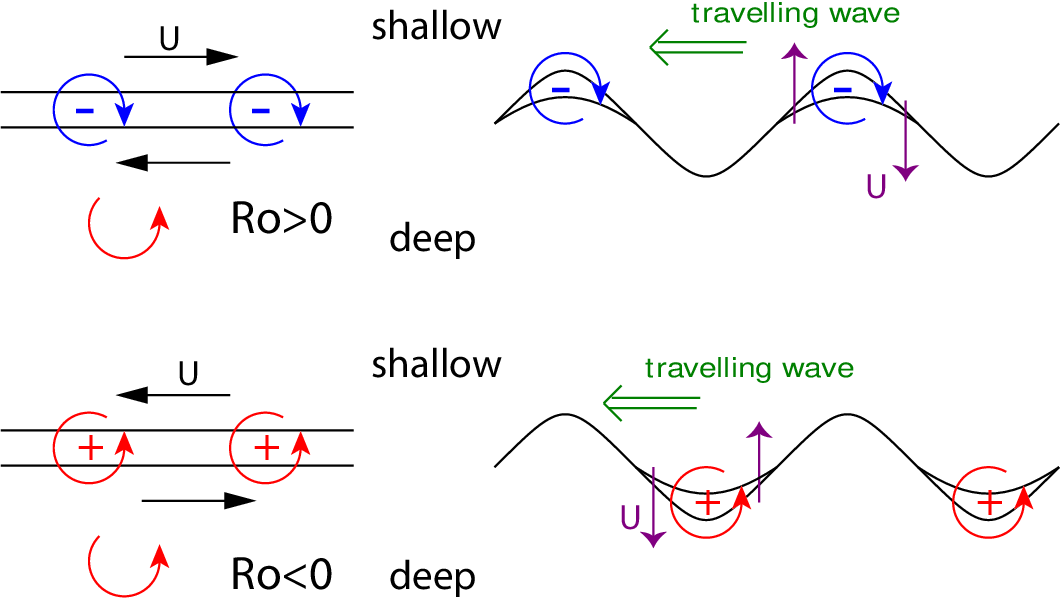}}
  \begin{picture}(0,0)
    \put(103,15){\textcolor[rgb]{1,0,0}{$\Omega$}}
    \put(103,88){\textcolor[rgb]{1,0,0}{$\Omega$}}
    \put(276,25){\textcolor[rgb]{1,0,0}{$\omega$}}
    \put(206,108){\textcolor[rgb]{0,0,1}{$\omega$}}
  \end{picture}
  \vspace{-4mm}
	\caption{Instability mechanism showing a different spatial behavior depending on the sign of the Rossby number}
	\label{fig:instab_mech}
\end{figure}

We can observe on figure \ref{fig:e6crit} that many features depend on the sign of the Rossby number, for $\beta \neq 0$.
The most striking one is the spatial behaviour of the split-sphere arrangement (fig. \ref{fig:e6crit}b) : for $\Ro>0$ the instability fills the
outer region ($r>r_0$) with spiralling arms, whereas for $\Ro<0$ the critical mode seems
to extend more in the inner region ($r<r_0$), leaving the outer region almost at rest.
The constant $\beta$ case (fig. \ref{fig:e6crit}a) also shows long range extension in the outer region for $\Ro>0$, but no spiralling.

To understand all these features, we must turn to the instability mechanism, which is
strongly dominated by the slope $\beta$.
Figure \ref{fig:instab_mech} sketches this mechanism.
For a column of local vorticity $\omega$ moved to a deeper region, 
the conservation of mass and total vorticity ($2\Omega_0 + \omega > 0$) leads to the following behaviour :
If $\omega > 0$ the local vorticity is amplified, and if
$\omega < 0$ it is damped.

Hence, for $\Ro>0$ the vorticity sheet is negative implying that the perturbation
is damped in the deeper region, whereas it is amplified in the
shallow region.
For $\Ro<0$, it is the opposite.
In both cases, the amplified vorticity will induce a flow which induces
a drift of the pattern in the prograde direction (as for any Rossby wave).

This fundamental asymmetry between positive and negative Rossby numbers partly explains
the results obtained by \cite{Hollerbach:1}. We will discuss an example in \S\ref{sec:exp_ss}.

\subsection{Radial scalings}

In figure \ref{fig:e6crit}abc the perturbation of the shear layer spreads
across the whole volume whereas in the flat container case (fig. \ref{fig:e6crit}d)
or the thermal convection instability in a sphere \cite[see][]{Dormy:TC}, it is radially localized.
This is explained by a simple model detailed in appendix \ref{sec:RadScale}, and the
main results obtained are summarized in the following lines.
 
For small $\beta$ as in the inner part of the split-sphere arrangement, 
the critical mode has a radial extension comparable to its
azimuthal scale, without oscillations;
but for large $\beta$ as it is the case for the outer part of the split-sphere arrangement,
the instability consists of a rapid oscillation of radial length scale $\lambda$ related to the azimuthal wave-number $m$ by
\begin{equation}
\label{eq:oscil}
	\lambda \sim m \sqrt{\abs{2\beta}}
\end{equation}
This rapid oscillation is contained in an envelope of radial size
\begin{equation}
\label{eq:rad_scale}
  \Lambda \sim \pa{2\abs{\beta}}^{-1/2} \alpha^{-1} E^0
\end{equation}
Equation \ref{eq:rad_scale} implies that the radial extent of the perturbation is
independent of $E$.
This spatial structure corresponds to the spiralling of the Rossby wave observed
for variable $\beta$ (see fig. \ref{fig:e6crit}bc).
We may also notice a fundamental difference with the thermal convection case
investigated by \cite{Yano} and \cite{Jones} where the $E^{1/6}$ radial extension 
decreases with $E$, so that the instability remains localized for small enough $E$.

\subsubsection{Numerical results}

\begin{figure}
  \centerline{ \epsfxsize 10cm \epsffile{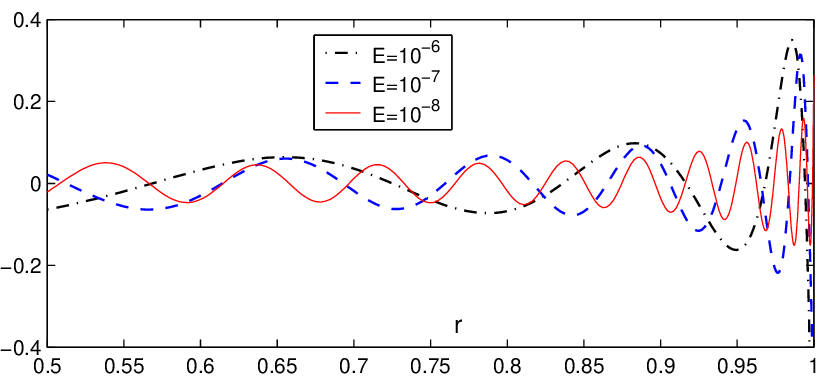}}
	\caption{Vorticity profile at fixed $\phi$ at
	the stability threshold for $\Ro > 0$ in the outer region.}
	\label{fig:rad_size_vort}
\end{figure}

We show in figure \ref{fig:rad_size_vort} a radial vorticity profile (at a fixed $\phi$) obtained with the split-sphere setup.
It shows that the vorticity gets amplified by the decreasing depth as approaching
the equator, while the spatial frequency of the oscillations increases,
in agreement with equation \ref{eq:oscil}.
In addition even when decreasing $E$, the perturbation seems
to spread across the whole width of the tank, as predicted by the scaling
\ref{eq:rad_scale}. This is not due to the singularity of the QG-model at the equator,
because we also observed the same phenomena with a truncated sphere, where the slope remains finite (see \S\ref{sec:split-sphere}).

\begin{figure}
  \centerline{ \epsfxsize 8cm \epsffile{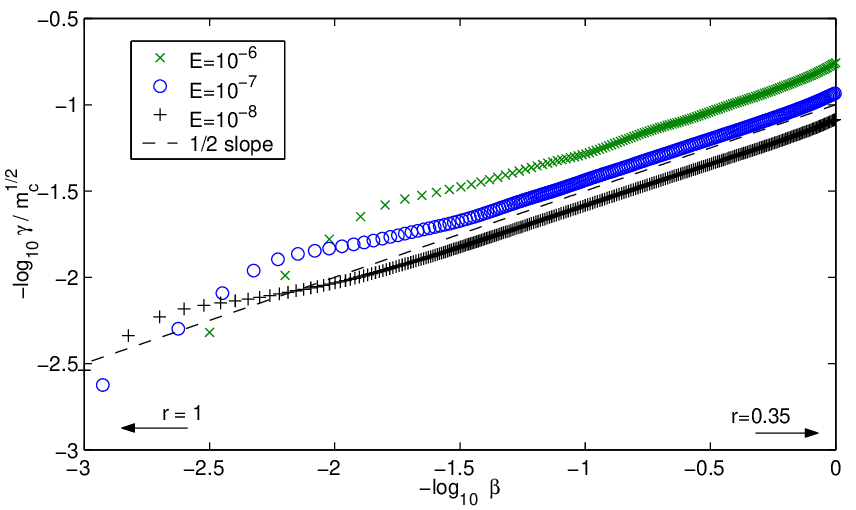}}
	\caption{Evolution of the radial size (evaluated using spatial
        derivation) of the unstable mode near the equatorial singularity
	for the split-sphere geometry.
	All these curves exhibit a wide range of slope $1/2$,
	which is the expected trend to cancel phase-mixing. Near the
        equatorial boundary ($r=1$) this scaling is no longer valid.}
	\label{fig:rad_size}
\end{figure}

The scaling \ref{eq:oscil} is observed quantitatively in numerical simulations
as shown on figure \ref{fig:rad_size}.
On this curves we can also see the effect of the equatorial boundary, which
modifies the scaling near $r=1$.
When $\beta$ is constant, there is no spiralling but the instability still fills
the whole gap, as figure \ref{fig:e6crit}a shows.

\section{Asymptotic laws for the onset}
\label{sec:Asymptotic}

\subsection{Scaling analysis}
\label{sec:scaling}

We perform a linear stability analysis on equation \ref{eq:Vort2D}.
Let $U(r)\, \vect{e_{\phi}}$ be the basic state velocity profile solution of the linear equation \ref{eq:baseflow} (for $\Ro=0$), and $\vect{u}$ the perturbation.
With $\Omega$ the vorticity associated with $U(r)$ :
$$
	\Omega = \Dp{U}{r} + \f{U}{r}
$$
we find
\begin{equation}
\label{eq:VortLin}
        \Dpt{\omega} + \Ro \pa{ u_r\Dp{\Omega}{r} + \f{U}{r} \Dp{\omega}{\phi} }
        - 2 \mean{\Dp{u_z}{z}} = E \lap \omega
\end{equation}
where we have neglected the small local vorticity $\Ro\:\Omega$.
We now introduce the frequency $f$ of the disturbance, and its
azimuthal wave-number $m$.

At the stability threshold, we expect the inertial terms to be both of the same magnitude, $$
		\Delta^{-2} \sim m^2
$$
which implies the wave-number of the perturbation to be comparable to the thickness of the shear layer
$$
		m \sim \Delta^{-1} \sim E^{-1/4}
$$

\subsubsection{Without slope}

When there is no slope equation \ref{eq:duzPump} reduces to
\begin{equation}
	\mean{\Dp{u_z}{z}} = -\f{E^{1/2}}{2L} \: \omega
\end{equation}
so that equation \ref{eq:VortLin} gives in terms of order of magnitude :
$$
        m f + \Ro \: \Delta^{-2} + \Ro \: m^2
        \sim \f{E^{1/2}}{L} m + E m^3
$$
With $m \sim E^{-1/4}$, the stability threshold is given
by a balance between the non-linear forcing and the viscous damping :
$$
		\Ro_c \sim E^{3/4}
$$
which is the result obtained by the linear theory of \cite{Busse:68} and by the critical local Reynolds number ($Re \sim \Ro E^{-3/4}$) theory of \cite{Niino}. It is also consistent with the experimental results of \cite{Niino} and \cite{Read} and corresponds to a Kelvin-Helmoltz instability of the shear layer.

For an oscillatory instability, the frequency $f$ should be consistent with
$m f \sim \Ro \: m^2$, and so
$$
	f \sim E^{1/2}
$$
This effect, if present, may be very difficult to separate from the advection of the pattern, because they are of similar amplitudes.

\subsubsection{With slope}
\label{sec:slope_scale}

When the slope is important, $\Dp{u_z}{z} \simeq \beta u_r$, and so
equation \ref{eq:VortLin} gives the following balance :
$$
        m f + \Ro \: \Delta^{-2} + \Ro \: m^2 + 2\beta
        \sim E m^3
$$
If again $m \sim E^{-1/4}$, the viscosity is negligible and the stabilization
is due to the slope : the instability has to overcome the \PT constraint.
As a matter of fact, from a balance between slope and forcing, we find
\begin{equation}
\label{eq:asymptotic_stab}
		\Ro_c \sim 2\beta E^{1/2}
\end{equation}
This is the expected scaling for non-flat containers such as spherical shells, and is quite
different from the flat case.
This result agrees with the criterion obtained by \cite{Kuo:49} applied to a Stewartson layer.

For a time-dependent instability, the frequency $f$ verifies $m f \sim 2\beta$,
and
$$
	f \sim \f{2\beta}{m}
$$
which is the Rossby wave dispersion relation.
At the onset, we may notice that the Ekman pumping is comparable to viscous dissipation in the bulk of the fluid.


\subsection{Rossby wave asymmetry effect}
\label{sec:Rayleigh}

\begin{figure}
  \centerline{ \epsfxsize 11cm \epsffile{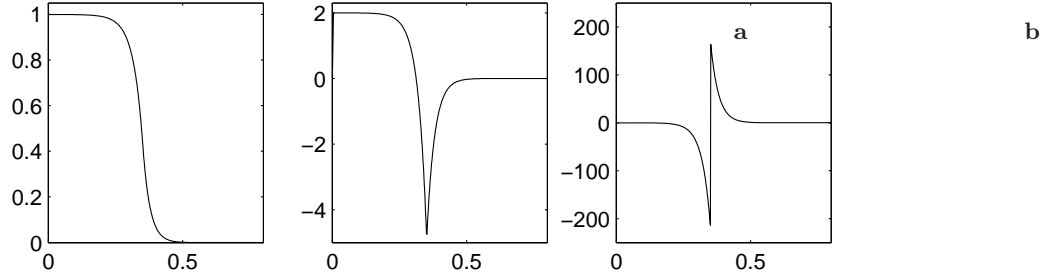}}
  \begin{picture}(0,0)(0,-10)
    \put(120,90){\textbf{a}}  \put(230,90){\textbf{b}}  \put(335,90){\textbf{c}}
  \end{picture}
  \vspace{-4mm}
	\caption{Basic state axisymmetric flow profiles (solution of eq. \ref{eq:baseflow}),
	  as a function of radius $r$
		for $E=10^{-6}$ and a split-sphere geometry with $r_0=0.35$.
		(a) rotation velocity $U/r$; (b) vorticity $\Omega$; (c) radial
		derivative of vorticity $\Dp{\Omega}{r}$.}
	\label{fig:base_profile}
\end{figure}
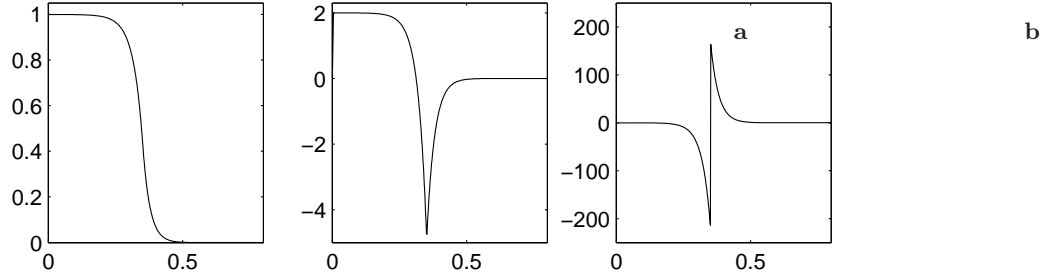

As stated in \S\ref{sec:sgnRo}, the instability develops as a Rossby wave.
Here we give a Rayleigh-like criterion for the onset of instability and use it to
evaluate the asymmetry induced by the Rossby wave.

The basic state flow $U$, represented on figure \ref{fig:base_profile} is the solution of the linear equation \ref{eq:baseflow}. The sign and amplitude of the forcing are both controlled by the parameter $\Ro$ in equation \ref{eq:VortLin}.
We are looking for solutions of the form
$\psi(r,\phi,t) = e^{im\phi+\lambda t} \psi(r)$.
We use the $\psi$ formulation introduced in section \ref{sec:scal_func},
but without the $\beta$-terms to keep simpler expressions.
Ignoring Ekman friction and viscosity, we obtain from equation \ref{eq:VortLin}
\begin{equation}
  \pa{\lambda + \f{im}{r}\Ro U} \omega\left[\psi\right]
     = \pa{2\beta - \Ro\Dp{\Omega}{r}}\f{im}{r} \psi
\end{equation}
Multiplying by $\psi^* L(r)r$ and integrating over $r$ we have
\begin{equation}
  \Int{0}{1} \omega\left[\psi\right] \psi^* Lr \:\dd r =
  \Int{0}{1} \f{ 2\beta - \Ro\Dp{\Omega}{r}}
               { \lambda + \f{im}{r}\Ro U } \: im \abs{\psi}^2 L\:\dd r
\end{equation}
Expanding the left hand side of this equation, and integrating by part, we obtain
\begin{equation}
  \Int{0}{1} \psi^* \pa{\f{m^2}{r^2}\psi-\f{1}{r}\Dp{}{r}\pa{r\Dp{\psi}{r}}} Lr\:\dd r
= \Int{0}{1} \f{m^2}{r^2}\psi \psi^* Lr\:\dd r + \left[\psi^*\Dp{\psi}{r}\right]_0^1 
- \Int{0}{1} \abs{\Dp{\psi}{r}}^2 Lr \:\dd r \nonumber
\end{equation}
and from the boundary conditions the bracketed term vanishes, leading to
\begin{equation}
  \Int{0}{1} \pa{\f{m^2}{r^2} \abs{\psi}^2 + \abs{\Dp{\psi}{r}}^2} Lr\:\dd r =
\Int{0}{1} \f{ im \abs{\psi}^2 \pa{2\beta - \Ro \Dp{\Omega}{r}} \pa{
\lambda^*-\f{im}{r}\Ro U}}
               {\abs{\lambda + \f{im}{r}\Ro U }^2} \:L\:\dd r \nonumber
\end{equation}
With $\Re(\lambda)$ the real part of the complex number $\lambda$,
the imaginary part of this equation gives
\begin{equation}
  0 = \Re(\lambda^*) \Int{0}{1} 
\f{ \abs{\psi}^2 \pa{2\beta - \Ro \Dp{\Omega}{r}} }
 {\abs{\lambda + \f{im}{r}\Ro U }^2} \:L\:\dd r
\end{equation}

For a mode to be unstable, we need $\Re(\lambda^*) \neq 0$, which means that
there is an $r$ for which 
\begin{equation}
\label{eq:Ro_opt}
  2\beta - \Ro\Dp{\Omega}{r} = 0
\end{equation}
This is exactly the criterion already obtained by \cite{Kuo:49}.
Using the complete pseudo-stream function given in section \ref{sec:scal_func} would lead to the same result. Therefore, the determination of the stability threshold will not be affected by using the small slope approximation, as it is usually done for the thermal convection case
\cite[see][]{Busse:70}.

\medskip


The critical Rossby number $\Ro_c$ that will satisfy \ref{eq:Ro_opt} will be
obtained for the maximum value $M$ of $\abs{\Dp{\Omega}{r}}$.
We must now take care of what value of $\beta$ comes into that expression.
Actually, the shear layer samples the values of $\beta$ about $\Delta$
to the left or to the right of the split radius $r_0$, depending on the sign of $\Ro$.
Then we have
\begin{equation}
\label{eq:Roc}
  \abs{\Ro_c^{\pm}} = \f{2\abs{\beta(r_0\pm\Delta)}}{M}
\end{equation}
At that point, remembering that the $E^{1/3}$-layer smoothes out discontinuities in $\Dp{\Omega}{r}$, we can expect that it will reduce the value of $M$, resulting in an higher $\Ro_c$.
Hence, the effect of the $E^{1/3}$-layer may be essentially to increase the critical Rossby number.


Assuming $\beta$ is a smooth function and $\Delta$ is small enough to use
a Taylor expansion
$$
\abs{\beta(r_0\pm\Delta)} = 
  \abs{\beta_0} \pm \Delta \pa{\Dp{\abs{\beta}}{r}}_0
$$
we obtain
$$
    \abs{\Ro^{\pm}} = 2 \, \Delta^2 \pa{ \abs{\beta_0} \pm
            \Delta \pa{\Dp{\abs{\beta}}{r}}_0 }
$$
This shows a fundamental asymmetry between positive and negative $\Ro$ due to the Rossby-wave nature of the instability.

\subsection{Numerical results}

\subsubsection{Split-sphere geometry}
\label{sec:split-sphere}

In the split-sphere geometry, the slope parameter $\beta$ becomes very large
near the equator and the expression of the Ekman pumping we use is no longer valid near the equator as the Ekman layer becomes singular (\cite{Roberts:63}; see also \S\ref{sec:Ekpump_valid}).
However, this area is very small and so we expect that it will not affect the
determination of the stability threshold.
To test this we followed the idea of \cite{Yano:nature} and truncated our container at
$r=0.9$ to prevent the slope from growing to large near the equator. The solutions showed the same features as the untruncated calculation and very similar stability thresholds.

\begin{figure}
  \centerline{ \epsfxsize 12cm \epsffile{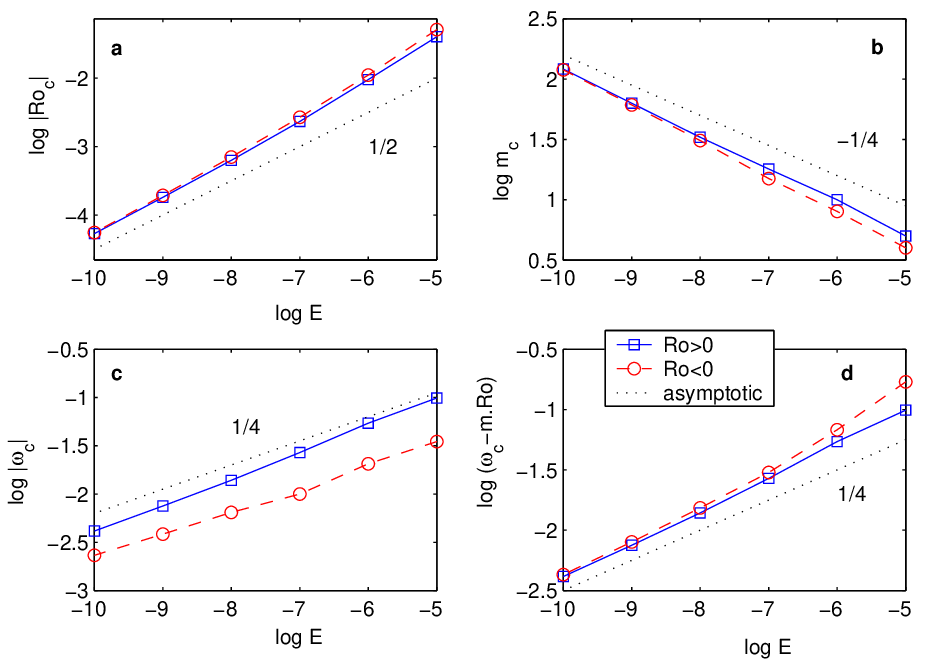}}
	\caption{Stability threshold data (given in table \ref{tab:stab}) as a function
	of the Ekman number.	(a) represents the threshold; (b) shows the critical wave
        number; (c) is the raw pulsation at the threshold; (d) is the pulsation corrected for advection effects. Dotted lines indicate the asymptotic exponent obtained in \S\ref{sec:slope_scale}.}
	\label{fig:stab}
\end{figure}

The results for the stability threshold are given in appendix \ref{tab:stab} and 
plotted in figure \ref{fig:stab}. They seem to match the asymptotical
scaling obtained previously, especially for small Ekman numbers.
We expect the perturbation to be a Rossby wave and in order to compare its pulsation with the
asymptotic laws, we need to correct for the advection by the mean flow.
Considering that the outer region (where the $\Ro>0$ instability develops) is not advected,
we don't correct the $\Ro>0$ data.
However the inner region (where the $\Ro<0$ case develops) can be considered as in solid body rotation (for small Ekman numbers)
and the simplest way to correct for the advection is then to subtract the rotation rate $\Delta\Omega$ from the phase speed $c=\omega/m$ for $\Ro<0$.
The pulsations of the Rossby waves are plotted in figure \ref{fig:stab}d,
and they agree quite well with the asymptotic law, for both positive and negative $\Ro$.

We checked that the difference between critical positive and negative Rossby number is proportional to $E^{1/4}$, in agreement with the findings of \S\ref{sec:Rayleigh}.

With our numerical code, we are able to test various boundary conditions. We report that
the boundary conditions (free-slip or no-slip) hardly affect the stability threshold,
whereas the presence of Ekman friction has a small yet visible effect.
The divergence correction we introduced in \S\ref{sec:scal_func}
seems to be negligible at the threshold.
The results of these tests are summarized in table \ref{tab:stab_fx}.

\subsubsection{Other geometries}
\label{sec:other_geom}

\begin{figure}
  \centerline{ \epsfxsize 6.5cm \epsffile{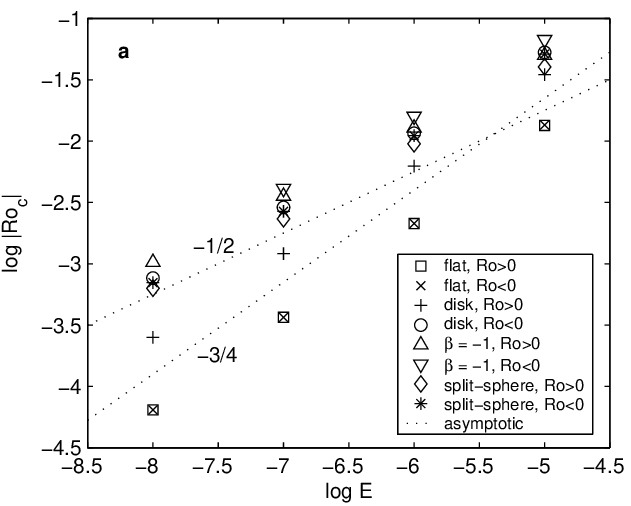}
  				\epsfxsize 6.5cm \epsffile{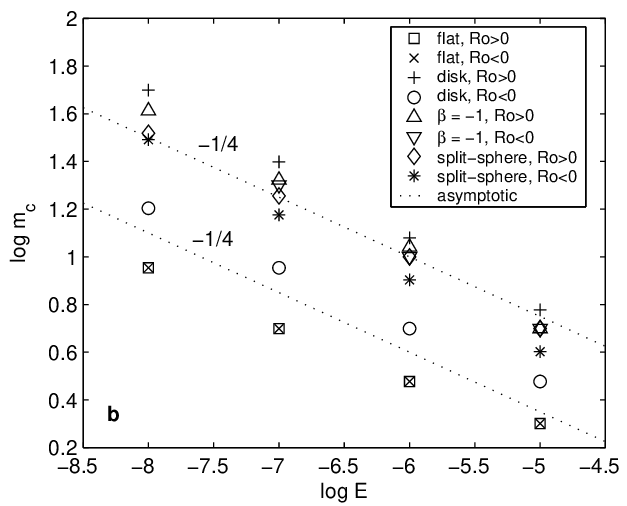}}
	\caption{Critical Rossby number (a) and critical wavenumber (b) for various geometries. Dotted lines indicate the asymptotic exponent obtained in \S\ref{sec:slope_scale}.}
	\label{fig:other_geom}
\end{figure}

We report the results of the numerical computations for the four studied geometries.
To compare the flat and non-flat case, we set up different geometries and do all the computation using our numerical implementation of the QG-model described in \S\ref{sec:2Dmodel}.
For the flat geometry we choose a flat cylindrical container with a height equal to its diameter. We also set up a container with an exponential shape, defined by $\beta=-1$ everywhere.
The split-sphere geometry studied in detail in \S\ref{sec:split-sphere} and finally
the disk setup.
This latter geometry is a split-sphere where the polar caps have been replaced
by flat disks to mimic our experimental setup (see fig. \ref{fig:exp_setup}).
However, with the QG-model we are not able to take the height-jump at the edge of the disks into account, so the resulting disk geometry is in fact flat for $r<r_0$ and spherical for $r>r_0$, with no height jump.
The results obtained are summarized in figure \ref{fig:other_geom}.
The critical wave-number data shows that the evolution of $m_c$ follows an
$E^{1/4}$ law, for both flat and non-flat cases, as expected from the asymptotic
study (\S\ref{sec:scaling}).
For the flat geometry, there is no difference between positive and negative Rossby number.
The split-sphere setup, the $\beta=-1$ case and the disk geometry for $Ro>0$
have comparable critical wave-numbers.

The flat geometry has much lower critical wave-numbers and Rossby numbers. From figure \ref{fig:e6crit}d we also see that the first unstable mode is totally symmetric when changing the sign of $\Ro$. It is also different from all the other cases (a,b and c) which present important slopes.
The disk geometry seem to be a hybrid case between flat and non-flat geometry :
From \ref{fig:e6crit}c, the spatial structure for $\Ro >0$ is very much like the
split-sphere setup, whereas for $Ro<0$ it seems to be somewhere in between split-sphere and flat geometry.
The figure \ref{fig:other_geom}b confirm this, with a critical wave-number between flat and non-flat. This is because the $Ro<0$ instability develops in the flat inner region.
On the other hand, figure \ref{fig:other_geom}a shows a critical Rossby number for $Ro<0$ very close to the split-sphere arrangement, whereas the $Ro>0$ seems to follow rather a slope-like $E^{3/4}$ law, even if the threshold is significantly higher than for the flat container.

\section{Comparison with experiments}
\label{sec:Exp}

\subsection{Flat container}

We reproduce numerically the constant depth experiments done by \cite{Read}.
The results of our linear calculation for the threshold are significantly
lower than the experimental ones, but are in good agreement with the theory of
\cite{Niino}.
The critical wave-length is in good agreement 
as shown in table \ref{tab:Read}.

To explain the discrepancy between experiment and theory,
\cite{Read} invoke the lack of the inner $E^{1/3}$ layer in the modeling of \cite{Niino},
which is also missing in this QG-model.

\begin{table}
$$
\begin{array}{ccccc}
      E              &  \Ro_c^F  &  m_c^F         & \Ro_c^S  & m_c^S   \\
      1.1 \, 10^{-5} &   0.044   &  6\textrm{-}8  & 0.021  & 6 \\
      8.3 \, 10^{-6} &   0.032   &  6\textrm{-}8  & 0.017  & 6 \\
      5.6 \, 10^{-6} &   0.025   &  6\textrm{-}8  & 0.013  & 7 \\
\end{array}
$$	
	\caption{Comparison between the experimental results of
	 \cite{Read} (denoted by $.^F$) and the numerical results we obtain
	 ($.^S$) for the stability threshold.}
	\label{tab:Read}

$$
\begin{array}{ccccc}
      E              &  \Ro_c^N   &  m_c^N         & \Ro_c^S  & m_c^S   \\
      3.0 \, 10^{-5} &   0.050    &  5\textrm{-}6  & 0.048    & 5 \\
      6.0 \, 10^{-6} &   0.015    &  7\textrm{-}9  & 0.0145   & 8 \\
\end{array}
$$	
	\caption{Comparison between the experimental results of
	 \cite{Niino} (denoted by $.^N$) and the numerical results we obtain
	 ($.^S$)for the stability threshold.}
	\label{tab:Niino}
\end{table}

\medskip

We also reproduce numerically the experiments of \cite{Niino}, which has the
particularity of having differential rotation only at the bottom of the tank, while
the top is rotating at the mean angular velocity.
Results are reported in table \ref{tab:Niino} and agree surprisingly well with
the original experimental data, as did their numerical calculations.

\subsection{Spherical shell geometry}
\label{sec:exp_ss}

The case of a spherical shell is quite singular : $\beta$ has opposite sign for $r<r_0$ and $r>r_0$, and the magnitude of $\beta$ is much higher for $r<r_0$.
This implies that Rossby waves travel in opposite directions in these two regions.
For positive $\Ro$, the instability may develop either in the $r>r_0$ region or in the $r<r_0$ one (see \S\ref{sec:sgnRo} and fig. \ref{fig:instab_mech}). In both directions the depth is decreasing, but from eq. \ref{eq:asymptotic_stab} we know that the instability will appear in the outer region ($r>r_0$) where the magnitude of $\beta$ is lower.
However, for negative $\Ro$ there is no deeper region around the shear layer, where the instability may be amplified by $\beta$-effect, so that we expect the threshold to be much higher and maybe the instability to be of a different kind, as the very small azimuthal wave-number observed by \cite{Hollerbach:1} may suggest.

\begin{figure}
  \centerline{ \epsfxsize 12cm \epsffile{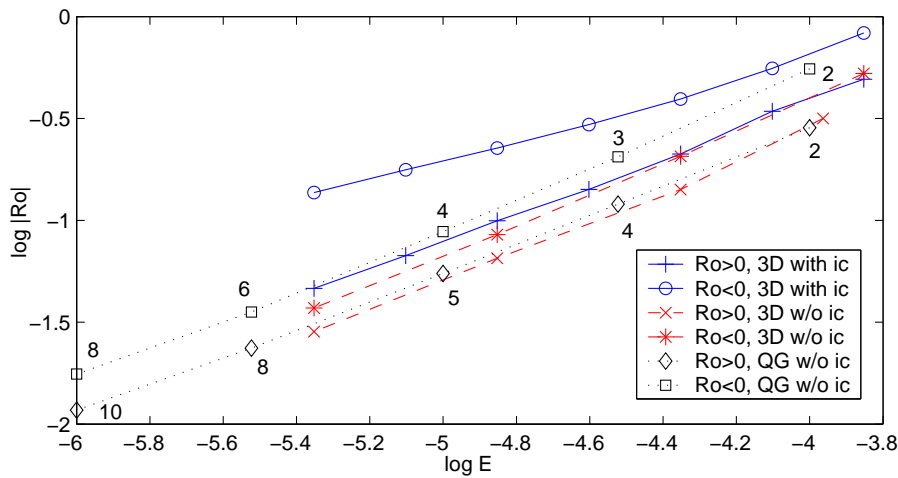}}
	\caption{Critical Rossby numbers obtained by \citeauthor{Hollerbach:1} (2003, summary of his fig. 4 and 8a) showing the impact of the slope on the stability threshold. We also plotted the corresponding results obtained with our QG-model (basic state shear layer obtained with an inner core, instabilities computed without) including the critical mode. When removing the inner core (ic), the threshold drops, and positive and negative Rossby numbers get closer. Our QG-model is close to the fully 3D results of Hollerbach}
	\label{fig:seuil_Hol}
\end{figure}

We reported on figure \ref{fig:seuil_Hol} the data found in his paper.
To check the effect of the geometry, he computed the basic-state shear layer in a spherical shell geometry, then took the profile obtained at mid-depth and extended it to all $z$, generating a $z$-invariant profile in a full sphere \cite[see][\S5]{Hollerbach:1}.
He finally computed the instabilities of that flow.
By artificially removing the inner sphere (and by that way loosing the singular geometrical property described above), \citeauthor{Hollerbach:1} recovers stability thresholds very close to the $\Ro>0$ case, which is much less modified by this process, for it is anyway controlled by the moderate slopes of the outer sphere.

We repeated exactly the same procedure with our QG-model : we computed the basic state shear layer with an inner-core ($r_0=0.35$), and then computed the stability threshold of this velocity profile in a full sphere geometry (without inner core). The results are compared with the 3D computations of \citeauthor{Hollerbach:1} on figure \ref{fig:seuil_Hol}. For $\Ro<0$, the QG-model gives thresholds about $30\%$ higher, and only $6\%$ higher for $\Ro>0$. The critical modes obtained are also in very good agreement.

This geometrical singularity also prevents any attempt to compute the instabilities of
the spherical shell setup with a QG-model.
We used the dynamo benchmark code described by \cite{DynBench} to perform a 
fully 3D numerical calculation at $E=10^{-4.5}$.
The critical Rossby number we find $\Ro_c = 0.16$ is in
very good agreement with the experimental one and with the numerical results of \cite{Hollerbach:1} as shown in figure \ref{fig:seuil_sphere}.


\subsection{Experimental study}

\subsubsection{Experimental setup}
\label{sec:ExpSetup}

\begin{figure}
  \centerline{ \epsfxsize 6cm \epsffile{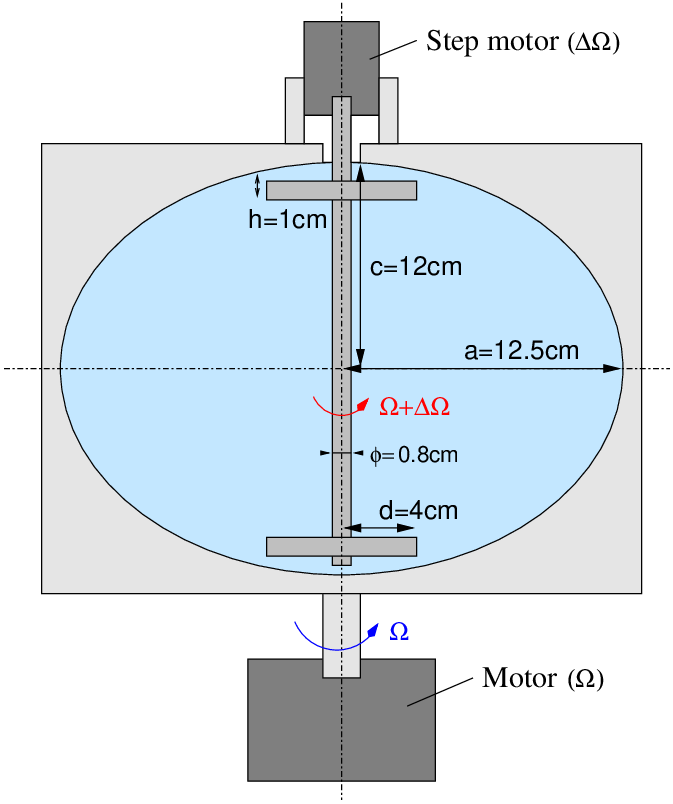}}
	\caption{Sketch of the experimental setup with disks geometry.
	Spherical shell geometry is obtained by replacing the disks by a
	sphere (4cm radius) on the inner rotating axis.}
	\label{fig:exp_setup}
\end{figure}

Figure \ref{fig:exp_setup} shows a sketch of the experimental set-up we use to study the onset of instability of the Stewartson layer. It consists of a plexiglass ellipsoid put in rotation by a brushless motor along its minor axis with a controlled angular velocity $\Omega_0$.
The working fluid is water. On the axis of rotation a shaft associated to a step motor can drive either an inner core or two disks.
Exact dimensions are reported in fig. \ref{fig:exp_setup} and further details can be found in \cite{Noir:03}.

\begin{figure}
  \centerline{ \epsfxsize 6.5cm \epsffile{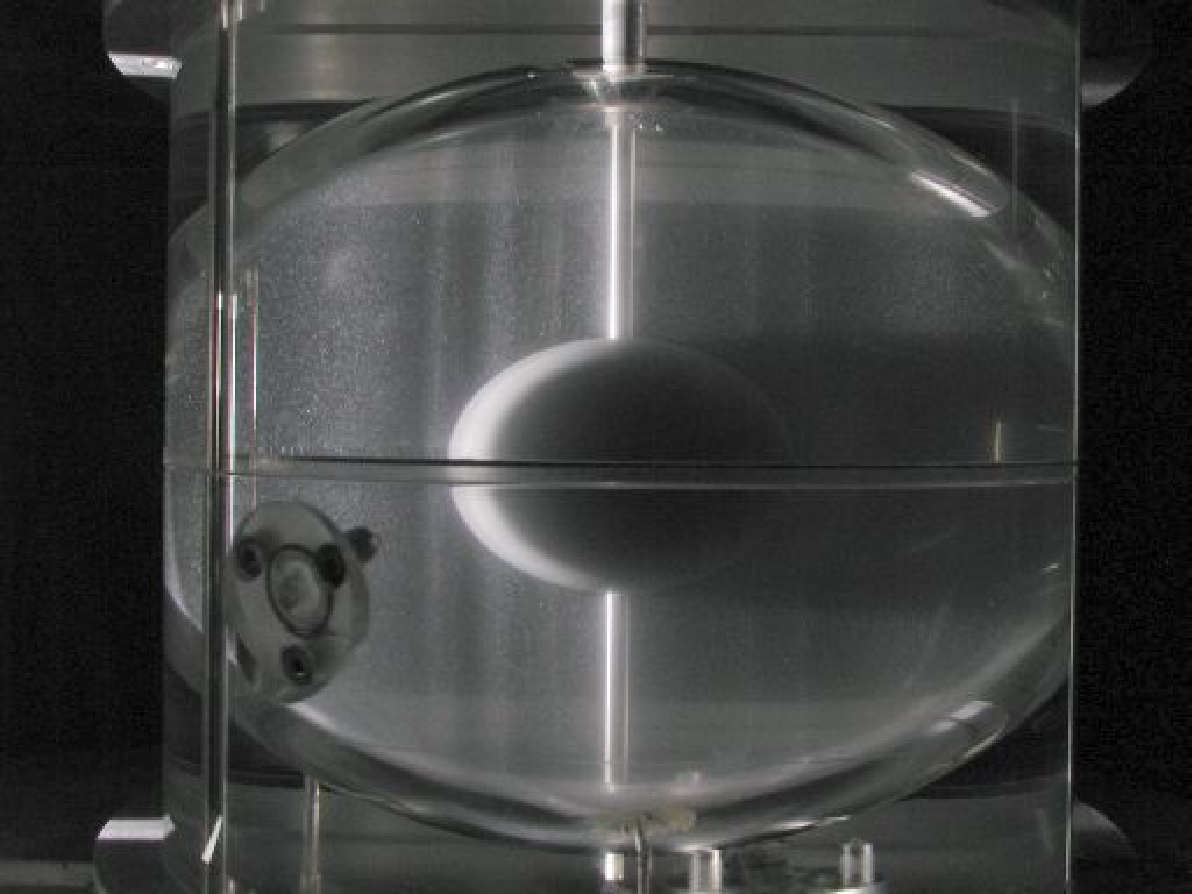}
	       \epsfxsize 6.5cm \epsffile{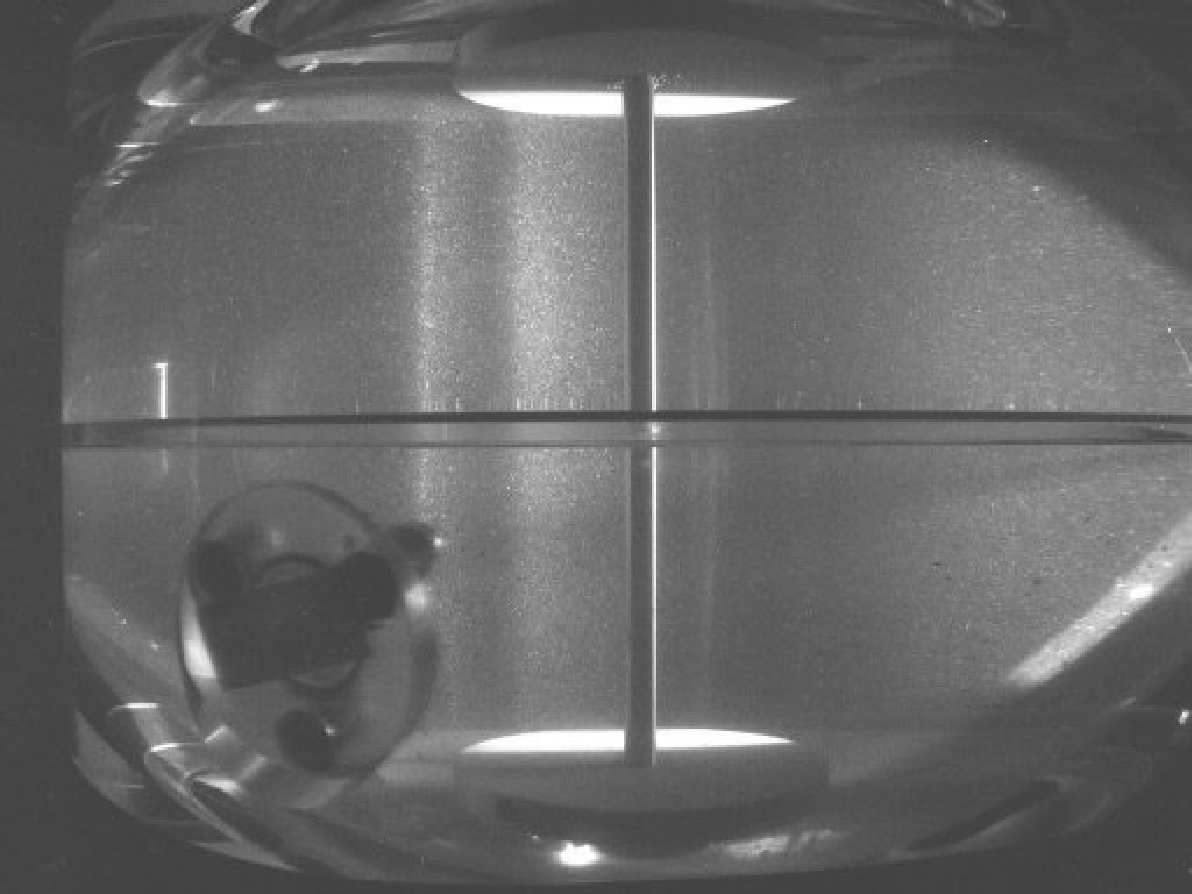}}
	\caption{Instabilities near the threshold as seen in our
	experiments (for $E \simeq 10^{-5}$). The vertical shaft is along the rotation axis.
	The features shown by the flakes are vertical, supporting the
	QG-approximation for our model.}
	\label{fig:DTO_exp}
\end{figure}

To determine the stability threshold, we use a flake solution (Kalliroscope
AQ 1000) and illuminate our container
with a light sheet in a plane including the rotation axis.
Variations of brightness allow us to visualize the shear-field in the illuminated plane.
When the shear-field is steady, we assume that the flow is stable, and when
the shear becomes unsteady (oscillations in time), the flow is unstable.
Two pictures of the observed patterns are shown on figure \ref{fig:DTO_exp}, showing
that the features of the flow are aligned with the rotation axis. The thin bright and
dark columns appear at the onset and are blinking, showing the advection of the pattern
of alternating vortices.
Our visual determination shown in figure \ref{fig:seuil_sphere} is in good agreement with
the numerical calculations of \cite{Hollerbach:1}. This makes us confident in the reliability of the optical method.

\subsubsection{Experimental results}

\begin{figure}
  \centerline{ \epsfxsize 12cm \epsffile{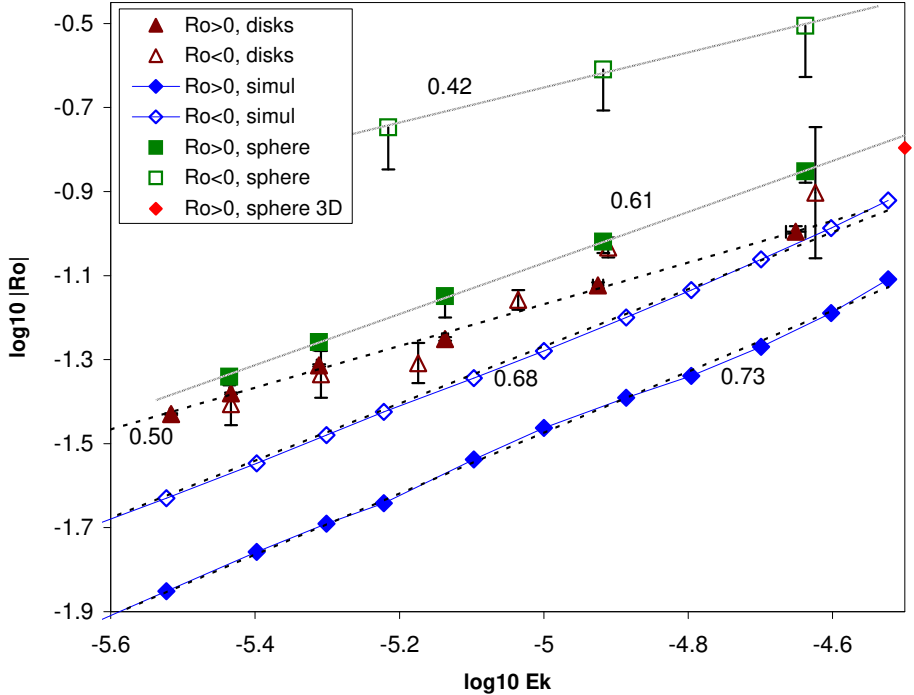}}
	\caption{Experimental determination of the stability threshold for two
        geometries, compared with the numerical calculations using a quasi-geostrophic model.
 \emph{sphere 3D} is computed with a fully 3D model. The negative differential rotation data are not well defined. The numbers are the slopes of the linear fits.}
	\label{fig:seuil_resume}
\end{figure}

The threshold is obtained for different values of the Ekman number with the
two geometries (inner core or two disks) for prograde differential rotation.
Results for the stability threshold are summarized in figure
\ref{fig:seuil_resume} and quantitative data are reported in tables \ref{tab:stab_sphere}
and \ref{tab:stab_disks}

Note that in the case of negative differential rotation, we were not able to determine accurately a stability threshold. In some experiments hysteresis appeared, correlated with the apparition of air bubbles in the tank. We have also observed very low frequency oscillation is some cases. We do not completely understand these results.
For the disks the data are not even on a straight line. This may be correct, as the numerical simulations also show some oscillations, due to the integer nature of the azimuthal wave number (see fig. \ref{fig:seuil_resume}).

The disk case seems to agree with the $E^{1/2}$ scaling, whereas the sphere case
is significantly steeper. However with only one decade of data, and for such high Ekman numbers, we do not expect the asymptotic laws to hold.

The numerical calculations for the disk model (see also \S\ref{sec:other_geom}), also shown in figure \ref{fig:seuil_resume} try to mimic the experimental setup : it is a split-sphere with flat polar caps with no height jump (we cannot model discontinuities in the height of the container).

The spherical-shell experiments are in agreement with the numerical calculations of \cite{Hollerbach:1}, as shown on figure \ref{fig:seuil_sphere}. The strong difference between positive and negative differential rotation is explained by the geometrical properties of this setup, as discussed above.

\section{Conclusion and Discussion}
\label{sec:ccl}

In this study, we give some insight into the instability of a
Stewartson shear layer in the very low Ekman regime.
The QG-model with the pseudo-stream function formalism is designed
for finite slopes as it handles the mass conservation correctly.
Despite failing to describe the $E^{1/3}$-layer, it has proved to be a
valuable tool for numerical experiments to study very low Ekman numbers.
It can also be used for non-linear calculations as long as the
quasi-geostrophic hypothesis stands, and may be thus useful for 
other ocean and planetary interior models.

At very low Ekman numbers $E < 10^{-6}$, we report an interesting radial structure,
comparable to the one exhibited in the thermal instability case by \cite{Dormy:TC}.
However, if the global structure of the flow is quite similar to
the spiralling thermal convection flow in a rapidly rotating spherical shell,
the radial extension of the instability is surprisingly independent of the Ekman number.

Our model allows us to compare flat top and bottom geometries with varying depth containers. Two different stabilizing processes are at work,
leading to different balances and scalings for the threshold : for flat containers the \cite{Busse:68} $E^{3/4}$ scaling applies, whereas with a slope a $\beta E^{1/2}$ scaling
is predicted. This is supported by numerical simulations for different geometries, and the following table summarizes the different cases for the instability
of a Stewartson layer  :

\begin{center}
\begin{tabular}{@{}cccccl}
\hline
geometry & $\Ro_c$ & sign of $\Ro$ & $m_c$ & $f_c$ & comment \\
\hline
	$\beta = 0$ & $E^{3/4}$ & $\Ro_c^{+} = -\Ro_c^{-}$ & $E^{-1/4}$ & $0$ ? & $Re_c = Cte$, Kelvin-Helmholtz-like\\
	constant $\beta \neq 0$ & $\beta E^{1/2}$ & $\Ro_c^+ \simeq -\Ro_c^-$ & $E^{-1/4}$ & $E^{1/4}$ & curvature effect, Rossby wave \\
	variable $\beta$ & $\beta E^{1/2}$ & $\Ro_c^+ \neq -\Ro_c^-$ & $E^{-1/4}$ & $E^{1/4}$ & spiralized Rossby wave \\
height jump & ? & not symmetric & $E^{-1/4}$ & ? &QG-model breaks down \\
\hline
\end{tabular}
\end{center}

\cite{Hollerbach:1} studied the difference between positive and negative
Rossby numbers, and pointed out that the geometry is the crucial feature,
through the radial derivative of the height of the container. The numerical
calculations he made shed some light on the dramatic changes in the critical mode
when switching from positive to negative Rossby numbers observed in some real experiments.
However, ignoring the Rossby wave processes at work in varying depth containers, he misunderstood his numerical results.

As we demonstrated in this paper, all the geometrical effects are due to a fundamental symmetry breaking in the Rossby wave propagation mechanism : depending on the sign of $\Ro$ and $\beta$, the instability will develop on one side of the shear layer or the other.
This symmetry breaking happens with any geometry, without the need for abrupt changes in the height of the container (even though this asymmetry will be magnified by such discontinuities).
Furthermore, the spherical shell setup is singular : when $\Ro<0$, the Rossby wave instability cannot develop on either side of the shear layer, leading to the singular results observed in this configuration.
Within this framework, the recent results of \cite{Hollerbach:1} may be interpreted correctly.

What still remains unclear is the nature of the instability in that particular case (spherical-shell and $\Ro<0$). Is it just a Rossby wave that needs much stronger forcing to grow, or another type of instability with significant 3D effects ?
This is even more annoying because this geometry is of geophysical interest for gaseous planets or planetary interiors. It is also the geometry used in the sodium experiment presented by \cite{Cardin02}, aimed at the study of induction processes in this type of flow, which are suspected to be able to drive a dynamo.
In addition, the non-linear and turbulent regimes have to be investigated, by means of numerical calculations and quantitative velocity measurement in experiments. It is clear that we need experiments at lower Ekman numbers, especially for spherical shells.



\section*{Acknowledgements}

Most of the computations presented in this paper were performed at the Service Commun de Calcul Intensif de l'Observatoire de Grenoble (SCCI). \\
We wish to thank Henri-Claude Nataf for stimulating discussions about this work.
We are also grateful to Natalia Bezaeva, Daniel Brito and Jean-Paul Masson
for their help in running the experiment. \\
This work was funded by the French Ministry of Research and
by the program \emph{Int\'erieur de la Terre} of the CNRS/INSU.

\appendix

\section{Ekman Pumping in a sphere}
\label{sec:EkP}


The Ekman pumping flow is given by \cite{Greenspan:book}
\begin{equation}
  \label{eq:EkP_vect}
  \left. \vect{u}.\vect{n} \right|_{\pm L} = -
     \f{E^{1/2}}{2} \vect{n}.\rot \paf{\vect{n}\vectoriel\vect{u} \pm \vect{u}}
     {\sqrt{\abs{\vect{n}.\vect{e_z}}}}
\end{equation}
with  $\vect{n}$ the \emph{outward} unit vector normal to the surface and
$\vect{u}$ the velocity jump between the geostrophic interior and the boundaries.
This Ekman pumping flow is the next order term in the $E^{1/2}$ development,
and is deduced from the quasi-geostrophic, $0$-order flow.
However, it has a dynamic action on that flow, and is thus important for
the dynamics of the system.
For simplicity we assume that the boundaries are at rest in the rotating frame.
It is straightforward to apply the result to moving boundaries by replacing
the velocity by the jump in velocity.

\subsection{Spherical coordinate system}

We can separate this
$$
  - \f{1}{2}E^{1/2}\pa{
    \underbrace{ \vect{n}.\rot
    \paf{\vect{n}\vectoriel\vect{u}}{\sqrt{\abs{\vect{n}.\vect{e_z}}}} }_{a}
    \pm \underbrace{
       \vect{n}.\rot\paf{\vect{u}}{\sqrt{\abs{\vect{n}.\vect{e_z}}}} }_{b} }
$$
We now use the spherical coordinate system $(\rho,\theta,\phi)$ 
to develop that expression, because it is the natural one for that boundary layer
($\vect{n} = \vect{e_\rho}$).
Term $a$ reduces to
$$
  a = \abs{\cos\theta}^{-1/2} \pa{ \pm \f{\abs{\tan\theta}}{2\rho} u_{\theta}
        - \Dp{u_\rho}{\rho} - \f{2}{\rho} u_\rho }
$$
and term $b$ can be written as
$$
  b = \abs{\cos\theta}^{-1/2} \pa{\pm \f{\abs{\tan\theta}}{2\rho} u_{\phi} + \omega_\rho }
$$
On the boundary, we have $\rho=1$, and so
\begin{equation}
  \label{eq:EkP_spheric}
  \left. \vect{u}.\vect{n} \right|_{\pm L} = \f{E^{1/2}}{2\abs{\cos\theta}^{1/2}} \pa{
  	\f{\abs{\tan\theta}}{2}\pa{\pm u_{\theta} - u_{\phi}} \: \pm \omega_\rho \:
  	- 2 u_\rho -\Dp{u_\rho}{\rho} }
\end{equation}

\subsection{QG-model flow}

We now assume the use of our QG approximation described in \S\ref{sec:2Dmodel}.
We need to translate \ref{eq:EkP_spheric} into cylindrical coordinate 
system $(r,\phi,z)$. We have
\begin{eqnarray*}
  \rho\sin\theta = r & \quad & \rho\cos\theta = z \\
  u_{\theta} = u_r \cos\theta - u_z \sin\theta & \quad & 
  u_\rho = u_r \sin\theta  + u_z \cos\theta
\end{eqnarray*}
$$
  \Dp{}{\rho} = \Dp{r}{\rho}\Dp{}{r} + \Dp{z}{\rho}\Dp{}{z} = 
     \sin\theta \Dp{}{r} + \cos\theta \Dp{}{z}
$$
Furthermore, $u_r$ and $u_{\phi}$ are $z$-independent, and $u_z$
is a linear function of $z$ (see section \ref{sec:BC}) :
\begin{equation}
\label{eq:uz_z}
  u_z = z \: \f{-r}{1-r^2} \, u_r(r,\phi)
\end{equation}
%
We find
\begin{equation}
  \Dp{u_\rho}{\rho} = \f{-2r}{1-r^2} u_r
\end{equation}

We also need to develop $\omega_\rho$ :
$$
  \omega_\rho = \omega_z \cos\theta + \omega_s \sin\theta
$$
where $\omega_r$ is given by
$$
  \omega_r = \f{1}{r}\Dp{u_z}{\phi} = \f{-z}{1-r^2} \Dp{u_r}{\phi}
$$
and finally
\begin{eqnarray}
  \left. \vect{u}.\vect{n} \right|_{\pm L} & = & \f{E^{1/2}}{2(1-r^2)^{1/4}} \left(
    \f{-r}{2(1-r^2)^{1/2}} u_{\phi} + \f{5}{2}\f{r}{1-r^2} u_r \right. \nonumber \\
    & & -\left. (1-r^2)^{1/2} \pa{\omega_z - \f{r}{1-r^2} \Dp{u_r}{\phi} } \right)
  \label{eq:EkP_2D}
\end{eqnarray}

This expression can be treated just like the $0$-order no-slip boundary condition :
$$
  \left. \vect{u}.\vect{n} \right|_{\pm L} = r u_r \pm \sqrt{1-r^2} u_z
$$
from which we deduce $u_z$ at the top and bottom boundary
$$
  \left. \pm u_z \right|_{\pm L} = 
  \pa{1-r^2}^{-1/2} \pa{ \left. \vect{u}.\vect{n} \right|_{\pm L} - r u_r }
$$
and from the linear $z$-dependence of $u_z$ we obtain
\begin{eqnarray}
\label{eq:EkP_sphere}
  \Dp{u_z}{z}
    & = & \f{E^{1/2}}{2(1-r^2)^{3/4}} \pa{ - \omega_z + \f{r}{1-r^2} \pa{ \Dp{u_r}{\phi}
    -\f{1}{2} u_{\phi} } - \f{5r}{2(1-r^2)^{3/2}} u_r } \nonumber \\
    & & + \f{-r}{1-r^2} u_r \\ [2mm]
    & = & E^{1/2}.P(u_r,u_{\phi},r) + \beta u_r \nonumber
  \label{eq:duz1}
\end{eqnarray}
It has to be emphasized that this results from a development in $E^{1/2}$,
which is no more valid when approaching the equator.

\subsection{Validity}
\label{sec:Ekpump_valid}

In the expression \ref{eq:duz1}, we have an order $E^{1/2}$ and order $1$ term.
The development is no more valid when these two terms are of the same
order of magnitude.
Setting $r=1-\epsilon$, we have $1-r^2 \sim 2\epsilon$.
Keeping the dominant terms in $1/\epsilon$, the limit of validity is given by
$$
  \f{E^{1/2}}{2(2\epsilon)^{3/4}} \f{5}{2} \f{1}{(2\epsilon)^{3/2}} u_r 
     \sim \f{1}{2\epsilon} u_r
$$
from which we get
\begin{equation}
\label{eq:Ek2_5}
  \epsilon \sim \f{1}{2} \paf{5}{4}^{4/5} E^{2/5}
\end{equation}
which shows the $E^{2/5}$ singularity on the equator \cite[see][]{Roberts:63}.
The angular extension is easily obtained from geometrical considerations :
$$
	s = \cos \delta\theta \sim 1-\delta\theta^2/2 \sim 1-\epsilon
$$
so that $\delta\theta = (2\epsilon)^{1/2}$ and
\begin{equation}
  \delta\theta \sim E^{1/5}
\end{equation}

\section{A simple model for the radial scaling}
\label{sec:RadScale}

\subsection{Approximations and assumptions}

In figure \ref{fig:e6crit}abc the perturbation of the shear layer spreads
across the whole volume whereas in the flat container case (fig. \ref{fig:e6crit}d)
or even the thermal convection case in a sphere \cite[see][]{Dormy:TC}, it is radially localized. In this section we derive a simple
heuristic model that predict all these properties \emph{outside the Stewartson shear layer}.

\medskip

To simplify our study, we make a few approximations :
\begin{itemize}
  \item we neglect the curvature;
  \item we drop the viscous dissipation term, but keep the Ekman friction;
  \item we assume small forcing, ignoring the local vorticity term.
\end{itemize}


\noindent 
The curvature may be negligible if the scales involved are small enough.
Looking at the results from section \ref{sec:scaling}, we expect the scales
to shrink while
lowering the Ekman number. Thus, we replace the angular variable $\phi$ by $y$.
Dropping the viscous term lowers the order of the differential equation and
makes it easier to handle. Numerical simulations with no viscous term
or an artificially weakened viscous term show that the structure stays
very similar. In addition, the threshold is expected to be only slightly
affected in the asymptotic regime. This is because the Ekman friction is also a dissipative term. Finally, we can neglect the local vorticity term because
in the linear stability case it is a higher order effect, as shown in section
\ref{sec:Rayleigh}. We end up with
\begin{equation}
\label{eq:lin_cart}
   \Dpt{\omega} + \Ro \pa{ U \Dp{\omega}{y} +  u_r\Dp{\Omega}{r} }
        = 2 \mean{\Dp{u_z}{z}}
\end{equation}
The expression of the Ekman friction involves several terms. We will only keep
the vorticity contribution, so that
\begin{equation}
\label{eq:dUz_simple}
   \mean{\Dp{u_z}{z}} \simeq \beta u_r - \alpha E^{1/2} \omega
\end{equation}
where $\alpha$ is the geometrical factor for the Ekman
friction term, and for a sphere $\alpha  = 1/2 (1-r^2)^{-3/4}$
(see eq. \ref{eq:EkP_sphere}).

We now assume a perturbation of the form
$$ \Psi(r,y,t) = \psi(r) e^{i\pa{ky-\omega t}} $$
and we obtain
\begin{equation}
  \pa{-i\omega + \Ro U ik + E^{1/2}\alpha}\pa{k^2 - \DDp{}{r}}\psi
     + \pa{\Ro \Dp{\Omega}{r} - 2 \beta} ik \psi = 0
\end{equation}
In a more convenient form, we can write
\begin{equation}
\label{eq:scale_eq}
  \DDp{\psi}{r} = k^2 \pa{1+ \f{1}{k} \f{ \pa{2\beta - \Ro\Dp{\Omega}{r}}
    \pa{\omega - \Ro Uk-iE^{1/2}\alpha}}{\pa{\omega - \Ro Uk}^2 + E\alpha^2}} \psi
\end{equation}

\subsection{Outer part, large $\beta$}

In the outer part the basic state velocity profile $U$ is vanishing, as well
as $\Dp{\Omega}{r}$. This simplifies the previously established relation
\ref{eq:scale_eq} :
\begin{equation}
\label{eq:scale_out}
  \DDp{\psi}{r} = k^2 \pa{1+ \f{2\beta}{k} \: \f{\omega - iE^{1/2}\alpha}
      {\omega^2 + E\alpha^2}} \psi
\end{equation}
In the case of small Ekman number, $E\alpha^2$ is negligible compared to
$\omega^2 \sim E^{1/2}$, but we have to keep the small imaginary part.
We now set
$$ \gamma = k\abs{1+\f{2\beta}{k\omega}}^{1/2} $$

If $\abs{2\beta} \gg 1$, we can rewrite \ref{eq:scale_out} as
\begin{equation}
  \DDp{\psi}{r} = - \gamma^2 \pa{1-\f{2i\beta E^{1/2}\alpha}
                                   {k\omega^2 + 2\beta\omega}} \psi
\end{equation}
and the solution for constant $\alpha$ and $\beta$ is of the form
$$
 \psi = \exp\pa{\pm \gamma \f{\beta E^{1/2}\alpha}{k\omega^2 + 2\beta\omega}r}
        \: \exp(\pm i\gamma r)
$$
This implies a rapid oscillation of typical length scale $\gamma$
\begin{equation}
\label{eq:oscil2}
	\gamma \sim k \sqrt{\abs{2\beta}}
\end{equation}
modulated by a slow exponential decay with typical length scale $\delta$ given by
\begin{equation}
\label{eq:envelope}
  \delta \sim \gamma^{-1} E^{-1/2} \alpha^{-1} \omega
\end{equation}
Inside this envelope, the radial length scale $\gamma^{-1}$ of the
perturbation decreases when $\beta$ increases. It corresponds to the spiralling
of the Rossby wave observed for variable $\beta$ (see fig. \ref{fig:e6crit}bc)

Using the scalings for the Stewartson layer instability $k \sim E^{-1/4}$
and $\omega \sim E^{1/4}$, relation \ref{eq:envelope} leads to
\begin{equation}
\label{eq:rad_scale2}
  \delta \sim \pa{2\abs{\beta}}^{-1/2} \alpha^{-1} E^0
\end{equation}
which implies that the radial extension of the perturbation is
independent of $E$.
This is a fundamental difference with the thermal convection case
investigated by \cite{Yano} and \cite{Jones} where the
scaling $k \sim E^{-1/3}$ and $\omega \sim E^{1/3}$ implies
\begin{equation}
  \delta \sim \pa{2\abs{\beta}}^{-1/2} \alpha^{-1} E^{1/6}
\end{equation}
which corresponds to a slowly decaying radial extension when decreasing $E$.

\subsection{Inner part, small $\beta$}

The inner part behaves exactly as the outer one, replacing $\omega$ by
$\omega - \Ro Uk$.
For a spherical shell it corresponds to the small $\beta$ case.

If $\abs{2\beta} < k\omega$, we can rewrite \ref{eq:scale_out} as
\begin{equation}
  \DDp{\psi}{r} = \gamma^2 \pa{1-\f{2i\beta E^{1/2}\alpha}
                                   {k\omega^2 + 2\beta\omega}} \psi
\end{equation}
which actually exchanges the length scale of the exponential decay and of the
oscillation compared to the outer region.
The solution for constant $\alpha$ and $\beta$ is then
$$
 \psi = \exp(\pm\gamma r) \: 
        \exp\pa{\mp i\gamma \f{\beta E^{1/2}\alpha}{k\omega^2 + 2\beta\omega}r}
$$
which is an exponential decay of characteristic length $\gamma^{-1}$,
inside which there are much larger length scale oscillations, which cannot be
detected.

For the Rossby wave, we have $\omega \sim k^{-1}$, so that
$ \gamma \sim k$, with a correction independent of $E$.
This shows that in the small
$\beta$ case the critical mode has a radial extension comparable to its
azimuthal scale.

\section{Stability threshold data}
\label{sec:StabData}

\subsection{Split-sphere stability threshold}
\label{tab:stab}

	Linear stability threshold values for different
	Ekman numbers $E$. $\textrm{NR}$ denotes the number of radial grid points
	used for the calculation. These values where obtained by QG numerical
	calculations with \emph{no-slip boundary conditions} and including
	\emph{Ekman friction}.

%

$$
\begin{array}{ccccc}       
        E          &    \Ro_c        & m_c &  \omega_c       & \textrm{NR} \\
        \hline
        10^{-5}    & 4.017 \,10^{-2}   & 5   & 9.886 \: 10^{-2}   & 300  \\
                   & -5.122 \,10^{-2}  & 4   & -3.495 \: 10^{-2}  &   \\
        10^{-6}    & 9.505 \,10^{-3}   & 10  & 5.428 \: 10^{-2}   & 400  \\
                   & -11.08 \,10^{-3}  & 8   & -2.060 \: 10^{-2} &   \\
        10^{-7}    & 2.326 \,10^{-3}  & 18  & 2.689 \: 10^{-2}  & 600   \\
                   & -2.674 \,10^{-3} & 15  & -0.999 \: 10^{-2} &    \\
        10^{-8}    & 6.314 \,10^{-4}  & 33  & 1.388 \: 10^{-3}  & 800  \\
                   & -7.018 \,10^{-4} & 31  & -6.446 \: 10^{-3} &    \\
        10^{-9}    & 1.821 \,10^{-4} & 63 & 7.505 \: 10^{-3}    & 1200 \\
                   & -1.938 \,10^{-4} & 61 & -3.844 \: 10^{-3}  &  \\
        10^{-10}   & 5.357 \,10^{-5} & 121 & 4.128 \: 10^{-3}   &  1500 \\
                   & -5.543 \,10^{-5} & 119 & -2.317 \: 10^{-3}   &
\end{array}
$$	
%

\subsection{Boundary conditions and viscosity dependence}
\label{tab:stab_fx}

	Comparison between $\Ro_c$ of different numerical models for the split-sphere geometry.
	Values are the $\Ro_c$ obtained by a model divided by
	the reference $\Ro_c$ (\emph{ns,ep}) for $\Ro>0$ at a given
 	Ekman number.
	The models are referenced by \emph{ns} and \emph{fs} respectively
	for no-slip and free-slip boundary conditions,
	\emph{ep} for the Ekman pumping, \emph{dv} for horizontal
	divergence and \emph{vs} for bulk viscosity; $\times 2$ stands for
        calculation with doubled number of radial grid points.
	A $0$ denotes the absence of this feature.

%

$$
\begin{array}{ccccccc}
 E      & ns,ep & ns,ep0 & fs,ep0 & ns,ep,dv0 & ns,ep,vs0 & ns,ep,\times 2 \\
\hline
10^{-6} & 1.0   & 0.86   & 0.87   & 0.995     & 0.64      & 0.999  \\
10^{-7} & 1.0   & 0.91   & 0.91   & 1.001     & 0.72      & 1.000 \\
10^{-8} & 1.0   & 0.94   & 0.94   & 1.002     & 0.78      & 1.000 \\
\end{array}
$$	
%

\subsection{Spherical shell experimental threshold}
\label{tab:stab_sphere}

 Stability threshold experimental data obtained for the
 spherical shell geometry, in water ($\nu = 10^{-6}\U{m^2/s}$).
 See \S\ref{sec:ExpSetup} for details about the experimental setup.

$$
\begin{array}{ccc}
        E    &  \textrm{stable} \: \Ro & \textrm{unstable} \: \Ro \\
\hline
3.51\; 10^{-5}	& 0.175	& 0.183 \\
2.30\; 10^{-5}	& 0.135	& 0.141 \\
1.21\; 10^{-5}	& 9.29\; 10^{-2}	& 9.56\; 10^{-2} \\
7.30\; 10^{-6}	& 6.93\; 10^{-2}	& 7.10\; 10^{-2} \\
4.88\; 10^{-6}	& 5.30\; 10^{-2}	& 5.52\; 10^{-2} \\
3.67\; 10^{-6}	& 4.48\ 10^{-2}	& 4.56\; 10^{-2} \\
\hline
2.30\; 10^{-5}	& -0.302	& -0.313 \\
1.21\; 10^{-5}	& -0.232	& -0.246 \\
6.09\; 10^{-6}	& -0.165	& -0.179  \\

\end{array}
$$	
%

\subsection{Disks experimental threshold}
\label{tab:stab_disks}

	Stability threshold experimental data obtained for 
	the disks geometry, in water ($\nu = 10^{-6}\U{m^2/s}$).
	See \S\ref{sec:ExpSetup} for details about the experimental setup.

$$
\begin{array}{ccc}
       E    &  \textrm{stable} \: \Ro & \textrm{unstable} \: \Ro \\
\hline       
2.23\; 10^{-5}	& 10.1\; 10^{-2}	& 11.1\; 10^{-2} \\
1.19\; 10^{-5}	& 7.53\; 10^{-2}	& 8.06\; 10^{-2} \\
7.30\; 10^{-6}	& 5.61\; 10^{-2}	& 5.94\; 10^{-2} \\
4.88\; 10^{-6}	& 4.86\; 10^{-2}	& 5.08\; 10^{-2} \\
3.68\; 10^{-6}	& 4.17\; 10^{-2}	& 4.33\; 10^{-2} \\
3.04\; 10^{-6}	& 3.72\; 10^{-2}	& 3.99\; 10^{-2} \\
\hline
2.38\; 10^{-5}  & -12.5\; 10^{-2}	& -15.0\; 10^{-2} \\
1.23\; 10^{-5}	& -9.27\; 10^{-2}	& -9.53\; 10^{-2} \\
9.21\; 10^{-6}	& -6.95\; 10^{-2}	& -7.14\; 10^{-2} \\
6.70\; 10^{-6}	& -4.92\; 10^{-2}	& -5.20\; 10^{-2} \\
4.91\; 10^{-6}	& -4.63\; 10^{-2}	& -4.93\; 10^{-2} \\
3.68\; 10^{-6}	& -3.93\; 10^{-2}	& -4.00\; 10^{-2} \\

\end{array}
$$	
%

\end{document}